\documentclass[aps,prl,superscriptaddress,showkeys,showpacs,twocolumn,nofootinbib,notitlepage,10pt]{revtex4-1}
\usepackage{amsmath,amssymb,slashed}
\usepackage[utf8]{inputenc}
\usepackage{graphicx}
\usepackage{hyperref}
\usepackage{bm,bbold}
\usepackage{accents}
\usepackage{ulem}
\def\!{\mskip-\thinmuskip}
\newcommand{\di}{{\rm d}}

\newcommand{\e}{{\rm e}}

\newcommand{\du}{\uparrow\downarrow}
\newcommand{\llangle}{\langle\langle}
\newcommand{\rrangle}{\rangle\rangle}

\newcommand{\beq}{\begin{equation}}
\newcommand{\eeq}{\end{equation}}

\usepackage[usenames,dvipsnames,svgnames,table]{xcolor}

\begin{document}
%**************************************************************************************************
\title{Detecting anomalous CP violation in heavy ion collisions\\ through baryon-electric charge correlations}

\author{David Frenklakh}
\email[]{david.frenklakh@stonybrook.edu}
\affiliation{Center for Nuclear Theory, Department of Physics and Astronomy, Stony Brook University, Stony Brook, New York 11794-3800, USA}

\author{\mbox{Dmitri E. Kharzeev}}
\email[]{dmitri.kharzeev@stonybrook.edu}
\affiliation{Center for Nuclear Theory, Department of Physics and Astronomy, Stony Brook University, Stony Brook, New York 11794-3800, USA}
\affiliation{Co-design Center for Quantum Advantage, Department of Physics and Astronomy, Stony Brook University, Stony Brook, New York 11794-3800, USA}
\affiliation{Department of Physics, Brookhaven National Laboratory, Upton, New York 11973-5000, USA}

\author{Andrea Palermo}
\email[]{andrea.palermo@stonybrook.edu}
\affiliation{Center for Nuclear Theory, Department of Physics and Astronomy, Stony Brook University, Stony Brook, New York 11794-3800, USA}

%**************************************************************************************************
\begin{abstract}
The chiral magnetic effect (CME) and the chiral vortical effect (CVE)  induce a correlation between baryon and electric currents. We show that this correlation can be detected using a new observable: a mixed baryon-electric charge correlator. This correlator is proportional to the baryon asymmetry, suggesting a novel way to separate the chiral effects from the background in heavy ion collisions. 
\end{abstract}
\maketitle

\section{Introduction}
Local, event-by-event CP violation is expected to happen in heavy ion collisions due to topological transitions in QCD. These processes result in a local chiral imbalance, quantified by the chiral chemical potential $\mu_5$. As a consequence of the chiral anomaly~\cite{Adler:1969gk,Bell:1969ts} $\mu_5$ acts as a source of anomalous transport phenomena~\cite{Kharzeev:2004ey,Kharzeev:2007jp} (see ~\cite{Kharzeev:2015znc,Li:2020dwr} for reviews), such as the Chiral Magnetic Effect (CME) and the Chiral Vortical Effect (CVE) \cite{Fukushima:2008xe,Kharzeev:2007tn,Son:2009tf,Kharzeev:2010gr}. The CME occurs in the presence of a magnetic field $B$ and results in an electric current along the direction of the magnetic field. In the hot QCD medium with $N_f = 3$ light quark flavors, the CME current reads:
\beq \label{eq:CME_current}
\vec{j}_{CME}=\frac{2}{3}\ \frac{N_c\mu_5}{2\pi^2}\ e^2\vec{B},
\eeq
where $N_c=3$ is the number of colors and $e$ is the electron charge; the factor $2/3 = 4/9 + 1/9 +1/9$ results from summing over the charges of light quarks.

The CVE occurs in a chiral matter with non-vanishing vorticity $\omega$ and baryon chemical potential $\mu_B$. It results in a baryon current aligned with the angular velocity, which for $N_f=3$ reads:
\beq \label{eq:CVE_current}
\vec{j}^B_{\text{~CVE}} = \frac{N_c \mu_5  \mu_B }{\pi^2}\    \vec{\omega}.
\eeq
We stress that in the case of 3 light quark flavors the magnetic field induces electric current but no baryon current, and vorticity induces baryon current but no  electric current \cite{Kharzeev:2010gr}. 

The experimental search for these effects in heavy ion collisions involves analyzing angular correlations among charged hadrons or baryons. This is done through the so-called $\gamma$-correlators \cite{Voloshin:2004vk, STAR:2009wot, STAR:2009tro} that allow to reduce the background. However, a number of background effects, such as charge and momentum conservation and elliptic flow \cite{STAR:2022ahj,Wang:2009kd,Bzdak:2009fc,Schlichting:2010qia, Pratt:2010zn, Wu:2022fwz}, still contribute and dominate the measurements. This is why isolating the effects  of anomalous transport in heavy ion collisions is a major challenge. Specific techniques and experiments have been devised to this end, but only upper bounds to chiral effects have been identified so far \cite{STAR:2023ioo,STAR:2021mii,ALICE:2020siw,Xu:2023wcy,Xu:2023elq}. 

Here, we show explicitly how the value of $\gamma$-correlators, and particularly of the so-called $\Delta\gamma$ correlators is related to the magnitude of macroscopic currents \eqref{eq:CME_current}, \eqref{eq:CVE_current}. Using these results, we give a quantitative estimate of the magnitude of the chiral chemical potential required to explain the recent ALICE measurements of CVE and CME signals in PbPb collisions at 5.02 TeV \cite{Wang:2023xhn}. Our estimates are done in a simple analytical framework and are in line with other estimates of the chiral chemical potential in heavy ion collisions \cite{Yuan:2023skl,Yu:2014sla}.

Finally, we introduce a new mixed baryon-electric charge correlator. This correlator is sensitive to both CME and CVE and it is odd in the baryon asymmetry. Since the baryon asymmetry at mid rapidities can be controlled experimentally in the event-by-event analysis,   
this property of the proposed mixed correlator can allow to separate the anomalous transport signal from the background. 

\section{Chiral imbalance from the CVE signal}
The CVE current \eqref{eq:CVE_current} separates baryons along the direction of the angular velocity. Therefore, in a heavy ion collision we expect a \emph{baryon separation} perpendicular to the reaction plane. We denote the resulting baryon separation as $\Delta N_B^{\uparrow\downarrow}$:
\beq
\Delta N_B^{\uparrow\downarrow}=N_B^\uparrow + N_{\bar{B}}^{\downarrow} - N_B^{\downarrow} -N_{\bar{B}}^{\uparrow},
\eeq
where $N_{B(\bar{B})}^{\uparrow,\downarrow}$ is the number of (anti)baryons observed above (i.e. with azimuthal angle $\phi\in[0,\pi)$) and below ($\phi\in [\pi,2\pi)$) the reaction plane.

To obtain the baryon separation resulting from the CVE, we integrate eq.~\eqref{eq:CVE_current} over the reaction plane and over the entire history of the hydrodynamic stage (assuming that the pre-hydro contribution is relatively small). Choosing a reference frame such that the angular velocity points in the $y$-direction, we have:
\begin{equation} \label{eq:sep_via_muB}
\Delta N_B^{\du} = \int_{\tau_0}^{\tau_f} d\tau\,\tau\int dx\,d\eta\,\frac{N_c}{\pi^2} \mu_5\, \mu_B \, \omega,
\end{equation}
where $\tau$ is the proper time and $\eta$ is the space-time rapidity in Milne coordinates. The hydrodynamic evolution starts at $\tau_0 = 1\,$fm and $\tau_f$ is the freeze-out time. 

We consider a simplified setup, where the fluid is dominated by Bjorken-like flow ~\cite{Bjorken:1982qr,Florkowski:2010zz,Floerchinger:2015efa} and vorticity is a small correction on top of it. In this case, temperature and baryon chemical potential scale as $T=T_0\left(\tau_0/\tau\right)^{1/3}$ and $\mu_B ={\mu_B}_0\left(\tau_0/\tau\right)^{1/3}$, where the quantities with subscript ``$0$'' are evaluated at $\tau_0$. Furthermore, throughout the paper we will assume that the chiral chemical potential is time-independent, and all the quantities in eq.~\eqref{eq:sep_via_muB} are homogeneous in space. Under these assumptions, we get:
\begin{equation} \label{eq:baryon_separation}
\Delta N_B^{\du} = \frac{N_c}{\pi^2}\mu_5 L_x\Delta\eta \int_{\tau_0}^{\tau_f} d\tau\,\tau \mu_B \, \omega ,
\end{equation}
where $L_x= 2R - b$ is the width of the intersecting area of two identical circles of radius $R$ with centers at a distance $b$ (in the case of $^{208}$Pb nuclei, $R\simeq 1.2 A^{1/3}=7.1\,$fm). The rapidity interval $\Delta\eta$ will be specified in accordance with the experimental data. 

To compute the baryon separation it is necessary to know the time dependence of the vorticity, that has been addressed in several studies \cite{Jiang:2016woz,Deng:2016gyh,Karpenko:2021wdm,Huang:2020dtn,Ivanov:2017dff,Deng:2020ygd}. In this work we will adopt the fit proposed in ref.~\cite{Jiang:2016woz} for the rapidity interval $\eta\in[-1,1]$, based on the AMPT model \cite{Lin:2004en}. At $\sqrt{s_{NN}}=5.02$ TeV, it reads:
\beq
\omega(b,\tau) = ~A(b)+\e^{- \tau/\tau_R} \left(\frac{\tau}{\tau_R}\right)^{0.3} B(b),
\eeq
where $A(b) = ~1.28 \tanh(0.35 b)\cdot 10^{-3}$, $B(b)  = ~(1.1 + 0.52 b)\cdot 10^{-2}$, with $\tau_R\approx 1.5\,$fm and $b$ is the impact parameter in fm; $\omega$ is in fm$^{-1}$. 

The baryon separation averaged over many events is not directly observable in heavy ion collisions because the average $\mu_5$ vanishes. 
The studies of local CP violation thus employ the so-called $\gamma$-correlators, which we will now briefly review.

The azimuthal distribution of hadrons of type $i$ in a given event is parameterized as:
\begin{align} \label{eq:azimuthal_decomposition}
    \frac{dN_i}{d\phi_i}=&\frac{N_i}{2\pi}\left[ 1+ 2a \sin(\phi_i - \Psi_{RP})\right.\nonumber\\
    &\left. +\sum_k 2v_k \cos\left(k(\phi_i - \Psi_{RP})\right)\right],
\end{align}
where $N_i$ is the total multiplicity in the rapidity interval under consideration, $v_n$ are the flow coefficients and a non-vanishing coefficient $a$ accounts for a local CP violation. $\Psi_{RP}$ denotes the azimuthal angle of the reaction plane. The $\gamma$ correlator between two particle species $i$ and $j$ is defined as \cite{Voloshin:2004vk}:
\begin{equation}\label{eq: gamma definition}
    \gamma_{112}=\langle\cos(\phi_i+\phi_j-2\Psi_{RP})\rangle.
\end{equation}
where the angular bracket is defined as:
\begin{equation}
    \langle f(\phi)\rangle=\frac{1}{N}\int\di\phi \, \frac{\di N}{\di \phi} f(\phi).
\end{equation}
This correlator has been extensively used for the search of the CME, in which case $i$ and $j$ denote two species of charged hadrons. In CVE studies, similarly, $i$ and $j$ are two baryons.
A non-vanishing signal for $\gamma_{112}$, however, can be produced also by a number of effects that do not require a local CP violation. In addition to the directed flow, backgrounds due to momentum and charge conservation and resonance decays cannot be ignored \cite{STAR:2022ahj,Wang:2009kd,Bzdak:2009fc,Schlichting:2010qia, Pratt:2010zn, Wu:2022fwz}. To mitigate these effects, one can separately compute $\gamma_{112}$ between particles with (electric or baryon) charge of the same sign, $\gamma^{SS}$, and of the opposite sign, $\gamma^{OS}$, and define $\Delta\gamma = \gamma_{112}^{OS}-\gamma_{112}^{SS}$ \cite{STAR:2013ksd, STAR:2014uiw}.

The subtraction eliminates charge-independent backgrounds; however it has been realized that the experimentally measured $\Delta\gamma$ is largely dominated by the elliptic flow of resonances decaying into particles appearing in the correlator \cite{Voloshin:2004vk}. The component of $\Delta\gamma$ driven by chiral effects is only a fraction of the $\Delta\gamma$ observed experimentally; we denote this fraction as $f_{CE}$:
\begin{equation}\label{eq:cme fraction mult}
    \Delta\gamma_{CE} = f_{CE}\Delta\gamma^{\text{obs}.}_{CE}.
\end{equation}
The latest experimental estimates from the isobar run at $200\,$GeV by STAR give an upper bound for the CME fraction  $f_{CME}\lesssim 10\%$ ~\cite{STAR:2023ioo}, whereas in PbPb collisions at $5.02\,$TeV at the LHC the estimates vary but they are compatible with this bound \cite{CMS:2017lrw, ALICE:2020siw,ALICE:2022ljz}.

Focusing on $\Delta\gamma_{CE}$, one gets
\begin{align}\label{eq: delta gamma in terms of a}
\Delta\gamma_{CE} =4a_i a_j+a_i^2+a_j^2,
\end{align}
where we have used $a_i= -a_{\bar{i}}$ for particle $i$ and antiparticle ${\bar i}$, as dictated by charge conjugation. In the above formula we also consider the case where particle and antiparticle of the same species appear in the correlator, which leads to the term $a_i^2+a_j^2$. If this is not the case, that term should not be taken into account.

Using eq.~\eqref{eq:azimuthal_decomposition}, the baryon asymmetry can also be related to the parameter $a$. Only the protons and $\Lambda$ hyperons are observed, so\footnote{Without loss of generality, we set $\Psi_{RP}=0$.}:
\begin{align} \label{eq:asymm_via_a}
\Delta N^{\uparrow\downarrow}_B=& 2 \sum_{i=p,\Lambda}\left(\int_0^\pi\di\phi \frac{\di N_i}{\di \phi}-\int_\pi^{2\pi}\di\phi \frac{\di N_i}{\di \phi} \right)\nonumber\\
=&\frac{8}{\pi}(N_p a_p+N_\Lambda a_\Lambda),
\end{align}
where the factor 2 accounts for the antiparticles.

We will use this relation to get a quantitative estimate of $\mu_5$ at $5.02\,$TeV using the $\Delta\gamma_{CVE}$ recently extracted by the ALICE Collaboration \cite{Wang:2023xhn}. $\Lambda\bar{\Lambda}$ and $p\bar{p}$ correlations are not accounted for in the data, so $\Delta\gamma_{CVE}=4a_\Lambda a_p$, in accordance with the discussion after \eqref{eq: delta gamma in terms of a}. 

Both $\Delta \gamma_{CVE}$ and $\Delta N^{\uparrow\downarrow}$ depend on $a_p$ and $a_\Lambda$, but it is not possible to express one in terms of the other unless we introduce an additional equation. To this end, we assume flavor symmetry between $p$ and $\Lambda$, setting $a_p=a_\Lambda$. Furthermore, we use $N_\Lambda \approx 0.5\,N_p$. This ratio has been checked using the SMASH-vHLLE hybrid model~\cite{Schafer:2021csj}.
Under these assumptions:
\begin{equation}
    \left.\begin{matrix}
        &\Delta\gamma_{CVE} = 4a^2\\
        &\Delta N^{\uparrow\downarrow}_B=\frac{12N_p}{\pi}a
    \end{matrix}\right\}\quad \Rightarrow \Delta N^{\uparrow\downarrow}_B = \frac{6N_p}{\pi}\sqrt{\Delta \gamma_{CVE}}. \label{eq:asymm_via_dgamma_CVE}
\end{equation}
We can now estimate the chiral chemical potential needed to explain the CVE data. Using~\eqref{eq:baryon_separation}, this chiral chemical potential is:
\begin{align}
    \mu_5=\frac{6\pi N_p \sqrt{\Delta\gamma_{CVE}}}{N_c L_x\Delta\eta\int_{\tau_0}^{\tau_f} d\tau\, \tau\,\mu_B\omega} .
    \label{eq:mu5_CVE}
\end{align}

\begin{table*}[t] 
\begin{tabular}{c|c|c|c|c|c|c|c|c|c|c} 
    \text{~Centrality\;[\%]~~} & $~~b[\text{fm}]~~$ & $~~T_0[\text{MeV}]~~$  & $~~N_p~~$ & $~~\Delta\gamma_{CVE}^{\text{obs}.}~~$ & $~~\mu_5^{(CVE)}/T_0~~$ & $~~\Delta N_B~~$ & $~~eB_0[m_\pi^2]~~$ & 
$~~N_{\pi^+}~~$ &$~~\Delta\gamma_{CME}^\text{obs}.~~$ & $~~\mu_5^{(CME)}/T_0~~$\\ \hline
    $0-10$ & $3.5$ & $428$ & 34 & $8\cdot 10^{-5}$ & $2.3$ & $0.4$ & $1.8\cdot 10^{-2}$ & $770$ & $2\cdot 10^{-5}$ & $10.3$ \\
    $10-20$ & $6.0$ & $411$ & 24 & $2\cdot 10^{-4}$ & $3.3$ & $0.3$ & $3.8\cdot 10^{-2}$ & $520$ & $5\cdot 10^{-5}$ & $7.3$ \\
    $20-30$ & $7.8$ & $389$ & 16 & $3\cdot 10^{-4}$ & $4.0$ & $0.2$ & $5.0\cdot 10^{-2}$ & $355$ & $9\cdot 10^{-5}$ & $7.5$ \\
    $30-40$ & $9.2$ & $362$ & 11 & $6\cdot 10^{-4}$ & $6.1$ & $0.1$ & $6.7\cdot 10^{-2}$ & $230$ & $16\cdot 10^{-5}$ & $7.4$ \\
    $40-50$ & $10.5$ & $332$ & 7 & $12\cdot 10^{-4}$ & $9.2$ & $0.1$ & $7.4\cdot 10^{-2}$ & $145$ & $25\cdot 10^{-5}$ & $9.0$ \\
\end{tabular}
\caption{This table summarizes the estimates of the first two sections. Refer to the main text for the sources of each column. The values of $\Delta\gamma^{\text{obs.}}$ reported here are the results of experimental measurement, and the contribution coming from anomalous transport is obtained by multiplying them by $f_{CE}$ as shown in eq.\eqref{eq:cme fraction mult}.} \label{tab:master}
\end{table*}

The values of $\Delta\gamma_{CVE}^{\text{obs}.}$ as function of centrality are taken from the ref.~\cite{Wang:2023xhn}.
The CVE fraction, $f_{CVE}$, is still unknown, so we will relate it to the CME fraction. There are several estimates of the CME fraction in PbPb collisions at $5.02\,$TeV in the literature. We will use $f_{CME}= 10\%$ as the upper limit but note that there are also estimates much lower than $10\%$ \cite{ALICE:2020siw}, that would imply a value of $\mu_5$ significantly smaller than the one reported below. For the CVE, we expect that statistical fluctuations are much larger than for the CME due to the lower baryon multiplicity compared to inclusive charged hadrons (dominated by pions). We will assume that the background is statistical, and $f_{CVE}$ is suppressed w.r.t. $f_{CME}$ by a factor $\sqrt{N_\pi^2/N_pN_\Lambda}\approx30$ for $|\eta|<0.5$ at all centralities. Therefore, we estimate $f_{CVE}\approx f_{CME}/30=0.3\%$. 

The baryon chemical potential is modelled by a Bjorken flow such that its value at freeze-out is $\mu_B=1\,$MeV, in accordance with the recent ALICE data \cite{Ciacco:2023ekv,ALICE:2023ulv}. We assume the freeze-out to happen at $155$ MeV at all centralities. The initial temperature is taken from ref.~\cite{Zakharov:2021uza}, and the Bjorken scaling determines the final freeze-out time. The impact parameter at different centralities is taken from ref.~\cite{Loizides:2017ack}. The proton multiplicity is measured by ALICE in the rapidity interval $|\eta|<0.5$ in ref.~\cite{ALICE:2019hno}.

Table \ref{tab:master} summarises the data mentioned above and our findings, and also contains data on the CME which is described in the next section. The values of $\mu_5/T_0$ obtained are in line with the other estimates in the literature \cite{Yuan:2023skl,Yu:2014sla}. One can notice that the extracted value of $\mu_5/T_0$ increases in more peripheral collisions; however, this  may be a consequence of our simplified model and requires a further investigation. 

To conclude this section, we will link the CVE observable to another independent experimental observable: \emph{the baryon asymmetry}. The baryon asymmetry characterizes the overall excess of baryons over antibaryons in the presence of a positive baryon chemical potential $\mu_B$ (or vice versa in the case of negative $\mu_B$). 

Since the only baryons observed in a detector are protons, the observed baryon asymmetry is the same as the net proton number, $\Delta N_B=\Delta N_p = N_p - N_{\bar{p}}$. In a simplified version of the statistical hadronization model the ratio of antiproton to proton yields is determined by the baryon chemical potential: $N_{\bar{p}}/N_p = e^{-2\mu_B/T}$,
where the baryon chemical potential and temperature are taken at freezeout. For $\mu_B\ll T$ it implies:

\beq \label{eq:net_baryon}
\Delta N_B = \frac{2\mu_B}{T} N_p. 
\eeq
Note that the ratio $\mu_B/T$ is constant in Bjorken model, so $\Delta N_B$ is constant throughout the hydrodynamic evolution, in accordance with baryon number conservation. The values of the average $\Delta N_B$ as a function of centrality computed from \eqref{eq:net_baryon} are reported in table \ref{tab:master}.

Using eqs.~\eqref{eq:baryon_separation} and \eqref{eq:net_baryon} we can rewrite baryon separation in terms of the baryon asymmetry:

\beq \label{eq:baryon_separation in terms of asymmetry}
\Delta N_B^{\du} = \frac{\Delta N_B}{2 N_p}\frac{N_c}{\pi^2}\mu_5 L_x \Delta\eta \int_{\tau_0}^{\tau_f} d\tau\, \tau\, \omega(\tau)T(\tau).
\eeq

\section{Chiral imbalance from the CME signal}
To check the consistency of our framework we also estimate the chiral chemical potential from the data reported on the CME in the same system. The CME electric current \eqref{eq:CME_current} separates charges in a way similar to eq. \eqref{eq:CVE_current}. Following the previous section, and assuming that $B$ points in the $y$ direction, the number of charges separated by the CME current is\footnote{The number of charges is computed from $j_{CME}/e$.}:
\begin{align}\label{eq: charge separation}
    \Delta N^{\du}_Q=\frac{N_c\mu_5}{3\pi^2}L_{x}\Delta\eta\int \di \tau \, \tau eB. 
\end{align}
This equation is the analog of eq.~\eqref{eq:baryon_separation}. We assumed the magnetic field to be independent of $x$ and $\eta$. 

Once again, we relate this quantity to the $\Delta\gamma$ observable. In the case of $\Delta\gamma_{CME}$ the particles used in this correlator are charged hadrons, mostly pions and protons. 
The charge separation is obtained as in eq.~\eqref{eq:asymm_via_a}, and it reads:
\begin{align}\label{eq:charge sep with a}
    \Delta N^{\du}_Q = \frac{8}{\pi}(N_{\pi^+} a_{\pi^+}+ N_p a_p).
\end{align}
We notice here that, since the pion multiplicity is much larger than the proton one, the proton contribution to eq.~\eqref{eq:charge sep with a} can be ignored. 

In the case of the CME data, correlations between particles of the same species are taken into account, so from eq.~\eqref{eq: delta gamma in terms of a} we have $\Delta \gamma_{CME}=4a_\pi a_p+a_p^2 +a_\pi^2$. To express $\Delta N_Q^{\du}$ in terms of $\Delta \gamma$, we assume once again $a_{\pi}\simeq a_{p}=a$. Under this assumption:
\begin{align}
    \left.\begin{matrix}
        \Delta\gamma_{CME} =6a^2\\
        \Delta N^{\du}_Q = \frac{8}{\pi}N_{\pi^+} a
    \end{matrix}\right\}\quad \Rightarrow \quad \Delta N^{\du}_Q = \frac{8}{\pi\sqrt{6}}N_{\pi^+}\sqrt{\Delta\gamma_{CME}}  ,\label{eq:asymm_via_dgamma_CME}
\end{align}
and finally:
\begin{align}
\mu_5=\frac{24\pi N_{\pi^+}\sqrt{\Delta\gamma_{CME}}}{\sqrt{6}N_c L_x \Delta\eta\, \int_{\tau_0}^{\tau_f} \di\tau\, \tau eB }.
\label{eq:mu5_CME}
\end{align}

The estimate of $\mu_5$ from the above formula requires the knowledge of the time evolution of the magnetic field in the quark gluon plasma \cite{McLerran:2013hla,Yuan:2023skl,Gursoy:2018yai,Hattori:2016emy,Huang:2015oca}. An accurate understanding of this time evolution requires solving  resistive dissipative relativistic magnetohydrodynamics (see e.g. \cite{Dash:2022xkz}), which is very difficult. Instead, we have used the publicly available code from ref. ~\cite{Gursoy:2018yai}, that solves the Maxwell equations in a medium with constant electric conductivity $\sigma$ (without back-reaction). In particular, we use superMC~\cite{Shen:2014vra} to generate the spectator density distribution and the magnetic field generated by participants is neglected. 

The resulting magnetic field was parameterized at different values of impact parameter using a simple formula \cite{McLerran:2013hla}: 
$B= B_0(b)\left[1+\left(\tau/\tau_B\right)^2\right]^{-1}$. 
The initial magnetic field $B_0$ and the decay time $\tau_B$ are very sensitive to the value of electric conductivity of the plasma. We choose the value $\sigma=0.05\,$fm$^{-1}\approx 10\,$MeV that is consistent with lattice calculations \cite{McLerran:2013hla,Kaczmarek:2012mgs, Aarts:2007wj} for the characteristic temperature of the plasma produced at the LHC. This value leads to the decay time of magnetic field $\tau_B\approx 1.5$ fm at all values of centrality. 
The values of the initial magnetic field $B_0$ resulting from the simulation are provided in table \ref{tab:master}, along with the $\Delta\gamma_{CME}$ correlator for charge separation, from ref.~\cite{Wang:2023xhn}, and the pion multiplicity, from ref.~\cite{ALICE:2019hno}. 

In the last column of table \ref{tab:master} we report the values of the chiral chemical potential computed from eq.~\eqref{eq:mu5_CME} with the CME fraction set to $f_{CME}=10\%$. The values of chiral chemical potential we obtain from the CME measurements are reasonably close to the ones we have obtained from the CVE. This illustrates that both electric charge and baryon correlations can be consistently interpreted in terms of CME and CVE, at least within  the simple framework that we use.

\section{Mixed baryon-electric charge correlator}

With the goal of extracting the correlations between the anomalous electric and baryon currents predicted by the CME and the CVE, let us now introduce a  new mixed correlator that is affected by both of these chiral effects. 
As we have seen, the baryon separation given by eq.~\eqref{eq:baryon_separation in terms of asymmetry} is proportional to the baryon asymmetry, which can be separately measured event-by-event. However, the $\Delta\gamma_{CVE}$ correlator scales with $\Delta N_B^2$, which makes it challenging to observe this dependence experimentally. 

To isolate the anomalous effects, we thus propose the following mixed baryon-electric charge correlator:
\beq\label{eq: Gamma def}
\Gamma_{QB} =  \sum_{i,j} \llangle \cos(\phi_{Q,i} + \phi_{B,j} - 2\psi_{RP}) \rrangle,
\eeq
where $i$ runs over the species of electrically charged hadrons (Q) while $j$ runs over the baryon species (B). We have also introduced double angular brackets denoting a non-normalized expectation value:
\begin{equation}
    \llangle f(\phi)\rrangle =\int\di\phi \, \frac{\di N}{\di \phi} f(\phi).
\end{equation}

As we will show, the correlator \eqref{eq: Gamma def} is predicted by the CVE to be proportional to the baryon asymmetry $\Delta N_B$. The dependence on $\Delta N_B$ can be analysed on the event-by-event basis similarly to the study of chiral magnetic wave (CMW) \cite{Kharzeev:2010gd,Burnier:2011bf} by the STAR Collaboration ~\cite{Wang:2012qs,STAR:2015wza}, where events with different charge asymmetry were selected. This dependence can provide a clear signature of anomalous transport.

We use the framework introduced in the previous sections to study the correlator \eqref{eq: Gamma def}. Considering only pions, protons, and $\Lambda$ hyperons in the final state we have:
\begin{equation}
   \Gamma_{QB}=\sum_{\substack{
   i=\{\pi^{\pm},p,\bar{p}\}\\ 
   j=\{p,\bar{p},\Lambda,\bar{\Lambda}\}
   }} 
   \llangle \cos(\phi_{C,i} + \phi_{B,j} - 2\psi_{RP}) \rrangle.
\end{equation}

We now define a $\Delta \Gamma_{QB}$. Denoting with $B$ particles with baryon number $B=1$ and with $\bar{B}$ those with $B=-1$, the same-sign and opposite-sign correlators are defined as:
\begin{align}
    \Gamma^{SS}&=\Gamma_{+B}+\Gamma_{-\bar{B}}, \qquad
    \Gamma^{OS}&=\Gamma_{+\bar{B}}+\Gamma_{-B},
\end{align}
and 
\begin{equation}\label{eq:def Delta Gamma}
    \Delta\Gamma_{QB}=\Gamma^{OS}-\Gamma^{SS}.
\end{equation}

Using eq.~\eqref{eq:azimuthal_decomposition}, we can evaluate these quantities as before. Since the number of pions is much larger than that of the other particles at hand, we can neglect the terms that do not include $N_{\pi^+}$, which leads us to\footnote{The same formula can be obtained restricting the charged particle species to pions.}:
\begin{align}
    \Delta\Gamma_{QB}=4N_{\pi^+} a_{\pi^+}(N_\Lambda a_\Lambda+N_p a_{p}).
\end{align}
Now it can be realized that $\Delta\Gamma_{QB}$ is directly related
to $\Delta N_B^{\uparrow\downarrow}$ and $\Delta N_Q^{\uparrow\downarrow}$, the baryon and electric charge separations. Indeed, using eqs.~\eqref{eq:asymm_via_a} and ~\eqref{eq:charge sep with a}:
\begin{align}\label{eq:mixed corr and DeltaN}
\Delta\Gamma_{QB}=\left(\frac{\pi}{4}\right)^2\Delta N^{\uparrow\downarrow}_B\Delta N^{\uparrow\downarrow}_Q.
\end{align}
Notice that, in contrast to the previous sections, we do not need any additional assumptions on the values of $a_{\pi}$, $a_p$, and $a_\Lambda$ to obtain eq.~\eqref{eq:mixed corr and DeltaN}.

Using eqs. \eqref{eq:baryon_separation in terms of asymmetry} and \eqref{eq: charge separation}, the mixed correlator can also be expressed as:
\beq
\Delta\Gamma_{QB}=\Delta N_B\frac{\mu_5^2}{N_p}\frac{N_c^2}{96\pi^2}L_x^2\Delta\eta^2\int d\tau\,\tau\,T\,\omega \int \,d\tau'\,\tau'\,eB.
\eeq
This makes it clear that the new mixed correlator depends on $\mu_5^2$, same as $\Delta\gamma_{CME}$ and $\Delta\gamma_{CVE}$, so it does not vanish when averaged over many events. Most importantly, it depends linearly on the baryon asymmetry $\Delta N_B$.

\begin{figure}
    \centering
    \includegraphics[width=0.5\textwidth]{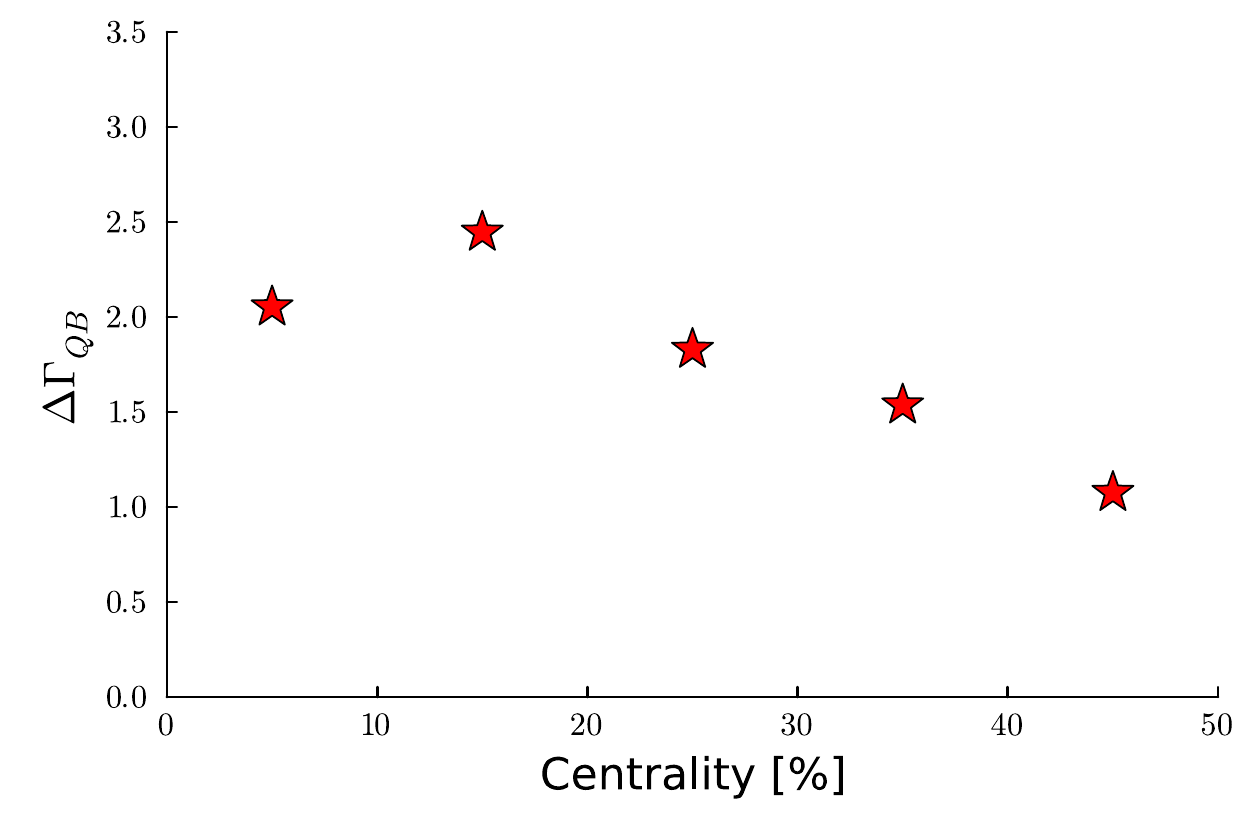}
    \caption{Prediction for the mixed correlator $\Delta\Gamma_{QB}$ (averaged over $\Delta N_B$) for PbPb collisions at $\sqrt{s_{NN}}=5.02\,$TeV based on eq.\eqref{eq:mixed_via_CVE_CME}.}
    \label{fig:mixed}
\end{figure}

The mixed correlator can also be expressed in terms of $\Delta\gamma_{CME}$, $\Delta\gamma_{CVE}$ and multiplicities. Using equations \eqref{eq:asymm_via_dgamma_CVE},\eqref{eq:asymm_via_dgamma_CME}:
\begin{align}
\Delta\Gamma_{QB}=\sqrt{\frac{3}{2}}N_p N_{\pi^+}\sqrt{\Delta\gamma_{CVE}\Delta\gamma_{CME}}.
\label{eq:mixed_via_CVE_CME}
\end{align}
Notice, however, that this relation is expected to hold only for the fraction of the correlators coming from anomalous transport, and deviations from it can help understand the backgrounds. In fig.~\ref{fig:mixed} we report our expectation of $\Delta\Gamma_{QB}$ based on eq.~\eqref{eq:mixed_via_CVE_CME} and using the values $\Delta\gamma^{\text{obs}.}$ reported in table \ref{tab:master}. Notice the non-monotonic behavior of the correlator with centrality, which results from the interplay of the decrease of multiplicities and the increase of the $\Delta\gamma$. 

\begin{figure}
    \centering
    \includegraphics[width=0.5\textwidth]{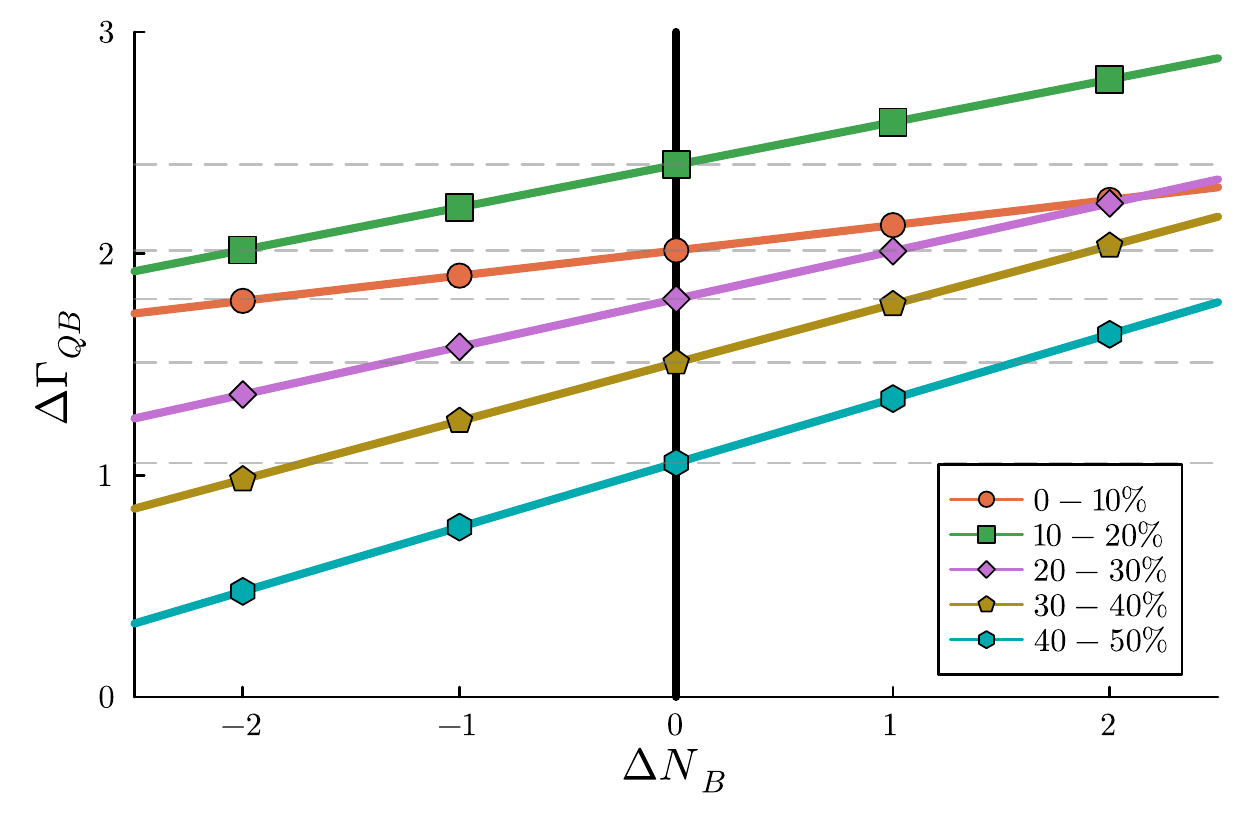}
    \caption{The predicted baryon asymmetry dependence of the mixed correlator $\Delta\Gamma_{QB}$ in different centrality bins. Dashed horizontal lines are plotted to guide the eye, and represents the $\Delta N_B$-independent background.} 
    \label{fig:gamma_vs_asym}
\end{figure}

\section{Anomalous dependence  on the event-by-event baryon asymmetry}
The most important feature of the correlator \eqref{eq:def Delta Gamma} is that the part of the signal induced by anomalous transport scales linearly with baryon asymmetry $\Delta N_B$. There is a priori no reason to expect a similar behavior from the background. 
From eq.~\eqref{eq:mixed corr and DeltaN}, we see that the anomalous part of $\Delta\Gamma_{QB}$  will change sign when $\Delta N_B$ changes sign. The baryon asymmetry is expected to be positive on average, due to the baryon stopping mechanism provided by the baryon junctions \cite{Kharzeev:1996sq, Brandenburg:2022hrp, Lv:2023fxk, ALICE:2013yba}, but it can have different value and sign event-by-event. Therefore, if the events are classified according to $\Delta N_B$, it should be possible to notice the linear scaling of $\Delta\Gamma_{QB}$ with $\Delta N_B$. A similar procedure was already employed in CMW studies, where the events where classified according to the value of the charge asymmetry \cite{Wang:2012qs,STAR:2015wza}. 

In the presence of background, we expect a relation
\begin{equation}
    \Delta\Gamma_{QB}(\Delta N_B) = f_\Gamma\ \frac{\Delta\Gamma_{QB}^*}{\Delta N_{B}^*} \Delta N_B +(1-f_\Gamma) \Delta\Gamma^*_{QB}, 
\end{equation}
where quantities with a star superscript denote a reference measurement of $\Delta\Gamma_{QB}$ at a specific $\Delta N_B^*$ and $f_\Gamma$ is the mixed-correlator signal fraction. 

To illustrate this behavior we use our estimate of the average value of $\Delta N_B$, as well as the $\Delta\gamma^{\text{obs.}}$ data reported in table~\ref{tab:master} to make a prediction for $\Delta\Gamma_{QB}(\Delta N_B)$. The value of $\Delta\Gamma^*$ is calculated using eq.~\eqref{eq:mixed_via_CVE_CME}. 
The signal fraction $f_\Gamma$ of the mixed correlator is assumed to be related to the CVE and the CME signal fractions through $f_\Gamma=\sqrt{f_{CME}\ f_{CVE}}$. Consistently with the previous sections, we use $f_{CME}\approx 10\%$ and $f_{CVE}\approx f_{CME}/30$, which leads to $f_\Gamma\approx 2\%$. Our estimates are reported in fig. \ref{fig:gamma_vs_asym}. One can see that the slope of the line is expected to slightly increase in peripheral collisions. The intercept is related to the $\Delta N_B$-even background and is non-monotonic as a function of centrality. 

\section{Conclusion}
The recent observation of baryon separation reported by the ALICE Collaboration \cite{Wang:2023xhn} at the LHC raises the question whether the signal is consistent with expectations from the CVE. In this paper, we have addressed this question by developing a model that allows to directly relate the magnitude of anomalous currents to the measured correlators. We find that the data can indeed be explained using a reasonable value of the chiral chemical potential. Moreover, the data on electric charge correlations can also be explained in the CME scenario by using a chiral chemical potential compatible with the one extracted from the baryon number correlations, again assuming the presence of CVE. 

Chiral anomaly predicts a correlation between the CME and the CVE \cite{Kharzeev:2010gr}, and this correlation can be used to isolate anomalous effects in heavy ion collisions. 
The interplay between the CME and the CVE also induces the mixing of the anomaly-driven collective excitations, the CMW and the Chiral Vortical Wave (CVW) \cite{Chernodub:2015gxa,Frenklakh:2016izv,Frenklakh:2017grl}. Recently, a particular mixed baryon-electric charge correlation has been computed in lattice QCD in the presence of a background magnetic field \cite{Ding:2023bft}.

We have introduced a new mixed electric-baryon charge correlator that can be used to detect anomalous transport in heavy ion collisions. The anomalous contribution to the mixed correlator is predicted to be directly proportional to the net baryon asymmetry. We thus propose to use such linear dependence of the mixed electric-baryon charge correlator on baryon asymmetry as a decisive test of the presence of local CP violation in heavy ion collisions.

\section{Acknowledgment}
We are grateful to  J. Liao, P. Tribedy and C. Wang for useful discussions. This work
was supported by the U.S. Department of Energy under Grants DE-FG88ER40388 and DE-SC0012704
(DK).

%%%%%%%%%%%%%%%%%%%%%%%%%%%%%%%%%%%%%%%%%%%%%%%%%%%%%%%%%%%%%%%%%%%%%%%%%%%%%%%%%%%%%%%%%%%%%%%%%%%%%
%                                       BIBLIOGRAPHY
%%%%%%%%%%%%%%%%%%%%%%%%%%%%%%%%%%%%%%%%%%%%%%%%%%%%%%%%%%%%%%%%%%%%%%%%%%%%%%%%%%%%%%%%%%%%%%%%%%%%%%
% \bibliographystyle{ieeetr}
% \bibliographystyle{unsrt}
\bibliography{biblio}

\begin{thebibliography}{67}%
\makeatletter
\providecommand \@ifxundefined [1]{%
 \@ifx{#1\undefined}
}%
\providecommand \@ifnum [1]{%
 \ifnum #1\expandafter \@firstoftwo
 \else \expandafter \@secondoftwo
 \fi
}%
\providecommand \@ifx [1]{%
 \ifx #1\expandafter \@firstoftwo
 \else \expandafter \@secondoftwo
 \fi
}%
\providecommand \natexlab [1]{#1}%
\providecommand \enquote  [1]{``#1''}%
\providecommand \bibnamefont  [1]{#1}%
\providecommand \bibfnamefont [1]{#1}%
\providecommand \citenamefont [1]{#1}%
\providecommand \href@noop [0]{\@secondoftwo}%
\providecommand \href [0]{\begingroup \@sanitize@url \@href}%
\providecommand \@href[1]{\@@startlink{#1}\@@href}%
\providecommand \@@href[1]{\endgroup#1\@@endlink}%
\providecommand \@sanitize@url [0]{\catcode `\\12\catcode `\$12\catcode
  `\&12\catcode `\#12\catcode `\^12\catcode `\_12\catcode `\%12\relax}%
\providecommand \@@startlink[1]{}%
\providecommand \@@endlink[0]{}%
\providecommand \url  [0]{\begingroup\@sanitize@url \@url }%
\providecommand \@url [1]{\endgroup\@href {#1}{\urlprefix }}%
\providecommand \urlprefix  [0]{URL }%
\providecommand \Eprint [0]{\href }%
\providecommand \doibase [0]{http://dx.doi.org/}%
\providecommand \selectlanguage [0]{\@gobble}%
\providecommand \bibinfo  [0]{\@secondoftwo}%
\providecommand \bibfield  [0]{\@secondoftwo}%
\providecommand \translation [1]{[#1]}%
\providecommand \BibitemOpen [0]{}%
\providecommand \bibitemStop [0]{}%
\providecommand \bibitemNoStop [0]{.\EOS\space}%
\providecommand \EOS [0]{\spacefactor3000\relax}%
\providecommand \BibitemShut  [1]{\csname bibitem#1\endcsname}%
\let\auto@bib@innerbib\@empty
%</preamble>
\bibitem [{\citenamefont {Adler}(1969)}]{Adler:1969gk}%
  \BibitemOpen
  \bibfield  {author} {\bibinfo {author} {\bibfnamefont {S.~L.}\ \bibnamefont
  {Adler}},\ }\href {\doibase 10.1103/PhysRev.177.2426} {\bibfield  {journal}
  {\bibinfo  {journal} {Phys. Rev.}\ }\textbf {\bibinfo {volume} {177}},\
  \bibinfo {pages} {2426} (\bibinfo {year} {1969})}\BibitemShut {NoStop}%
\bibitem [{\citenamefont {Bell}\ and\ \citenamefont
  {Jackiw}(1969)}]{Bell:1969ts}%
  \BibitemOpen
  \bibfield  {author} {\bibinfo {author} {\bibfnamefont {J.~S.}\ \bibnamefont
  {Bell}}\ and\ \bibinfo {author} {\bibfnamefont {R.}~\bibnamefont {Jackiw}},\
  }\href {\doibase 10.1007/BF02823296} {\bibfield  {journal} {\bibinfo
  {journal} {Nuovo Cim. A}\ }\textbf {\bibinfo {volume} {60}},\ \bibinfo
  {pages} {47} (\bibinfo {year} {1969})}\BibitemShut {NoStop}%
\bibitem [{\citenamefont {Kharzeev}(2006)}]{Kharzeev:2004ey}%
  \BibitemOpen
  \bibfield  {author} {\bibinfo {author} {\bibfnamefont {D.}~\bibnamefont
  {Kharzeev}},\ }\href {\doibase 10.1016/j.physletb.2005.11.075} {\bibfield
  {journal} {\bibinfo  {journal} {Phys. Lett. B}\ }\textbf {\bibinfo {volume}
  {633}},\ \bibinfo {pages} {260} (\bibinfo {year} {2006})},\ \Eprint
  {http://arxiv.org/abs/hep-ph/0406125} {arXiv:hep-ph/0406125} \BibitemShut
  {NoStop}%
\bibitem [{\citenamefont {Kharzeev}\ \emph {et~al.}(2008)\citenamefont
  {Kharzeev}, \citenamefont {McLerran},\ and\ \citenamefont
  {Warringa}}]{Kharzeev:2007jp}%
  \BibitemOpen
  \bibfield  {author} {\bibinfo {author} {\bibfnamefont {D.~E.}\ \bibnamefont
  {Kharzeev}}, \bibinfo {author} {\bibfnamefont {L.~D.}\ \bibnamefont
  {McLerran}}, \ and\ \bibinfo {author} {\bibfnamefont {H.~J.}\ \bibnamefont
  {Warringa}},\ }\href {\doibase 10.1016/j.nuclphysa.2008.02.298} {\bibfield
  {journal} {\bibinfo  {journal} {Nucl. Phys. A}\ }\textbf {\bibinfo {volume}
  {803}},\ \bibinfo {pages} {227} (\bibinfo {year} {2008})},\ \Eprint
  {http://arxiv.org/abs/0711.0950} {arXiv:0711.0950 [hep-ph]} \BibitemShut
  {NoStop}%
\bibitem [{\citenamefont {Kharzeev}\ \emph {et~al.}(2016)\citenamefont
  {Kharzeev}, \citenamefont {Liao}, \citenamefont {Voloshin},\ and\
  \citenamefont {Wang}}]{Kharzeev:2015znc}%
  \BibitemOpen
  \bibfield  {author} {\bibinfo {author} {\bibfnamefont {D.~E.}\ \bibnamefont
  {Kharzeev}}, \bibinfo {author} {\bibfnamefont {J.}~\bibnamefont {Liao}},
  \bibinfo {author} {\bibfnamefont {S.~A.}\ \bibnamefont {Voloshin}}, \ and\
  \bibinfo {author} {\bibfnamefont {G.}~\bibnamefont {Wang}},\ }\href {\doibase
  10.1016/j.ppnp.2016.01.001} {\bibfield  {journal} {\bibinfo  {journal} {Prog.
  Part. Nucl. Phys.}\ }\textbf {\bibinfo {volume} {88}},\ \bibinfo {pages} {1}
  (\bibinfo {year} {2016})},\ \Eprint {http://arxiv.org/abs/1511.04050}
  {arXiv:1511.04050 [hep-ph]} \BibitemShut {NoStop}%
\bibitem [{\citenamefont {Li}\ and\ \citenamefont {Wang}(2020)}]{Li:2020dwr}%
  \BibitemOpen
  \bibfield  {author} {\bibinfo {author} {\bibfnamefont {W.}~\bibnamefont
  {Li}}\ and\ \bibinfo {author} {\bibfnamefont {G.}~\bibnamefont {Wang}},\
  }\href {\doibase 10.1146/annurev-nucl-030220-065203} {\bibfield  {journal}
  {\bibinfo  {journal} {Ann. Rev. Nucl. Part. Sci.}\ }\textbf {\bibinfo
  {volume} {70}},\ \bibinfo {pages} {293} (\bibinfo {year} {2020})},\ \Eprint
  {http://arxiv.org/abs/2002.10397} {arXiv:2002.10397 [nucl-ex]} \BibitemShut
  {NoStop}%
\bibitem [{\citenamefont {Fukushima}\ \emph {et~al.}(2008)\citenamefont
  {Fukushima}, \citenamefont {Kharzeev},\ and\ \citenamefont
  {Warringa}}]{Fukushima:2008xe}%
  \BibitemOpen
  \bibfield  {author} {\bibinfo {author} {\bibfnamefont {K.}~\bibnamefont
  {Fukushima}}, \bibinfo {author} {\bibfnamefont {D.~E.}\ \bibnamefont
  {Kharzeev}}, \ and\ \bibinfo {author} {\bibfnamefont {H.~J.}\ \bibnamefont
  {Warringa}},\ }\href {\doibase 10.1103/PhysRevD.78.074033} {\bibfield
  {journal} {\bibinfo  {journal} {Phys. Rev. D}\ }\textbf {\bibinfo {volume}
  {78}},\ \bibinfo {pages} {074033} (\bibinfo {year} {2008})},\ \Eprint
  {http://arxiv.org/abs/0808.3382} {arXiv:0808.3382 [hep-ph]} \BibitemShut
  {NoStop}%
\bibitem [{\citenamefont {Kharzeev}\ and\ \citenamefont
  {Zhitnitsky}(2007)}]{Kharzeev:2007tn}%
  \BibitemOpen
  \bibfield  {author} {\bibinfo {author} {\bibfnamefont {D.}~\bibnamefont
  {Kharzeev}}\ and\ \bibinfo {author} {\bibfnamefont {A.}~\bibnamefont
  {Zhitnitsky}},\ }\href {\doibase 10.1016/j.nuclphysa.2007.10.001} {\bibfield
  {journal} {\bibinfo  {journal} {Nucl. Phys. A}\ }\textbf {\bibinfo {volume}
  {797}},\ \bibinfo {pages} {67} (\bibinfo {year} {2007})},\ \Eprint
  {http://arxiv.org/abs/0706.1026} {arXiv:0706.1026 [hep-ph]} \BibitemShut
  {NoStop}%
\bibitem [{\citenamefont {Son}\ and\ \citenamefont
  {Surowka}(2009)}]{Son:2009tf}%
  \BibitemOpen
  \bibfield  {author} {\bibinfo {author} {\bibfnamefont {D.~T.}\ \bibnamefont
  {Son}}\ and\ \bibinfo {author} {\bibfnamefont {P.}~\bibnamefont {Surowka}},\
  }\href {\doibase 10.1103/PhysRevLett.103.191601} {\bibfield  {journal}
  {\bibinfo  {journal} {Phys. Rev. Lett.}\ }\textbf {\bibinfo {volume} {103}},\
  \bibinfo {pages} {191601} (\bibinfo {year} {2009})},\ \Eprint
  {http://arxiv.org/abs/0906.5044} {arXiv:0906.5044 [hep-th]} \BibitemShut
  {NoStop}%
\bibitem [{\citenamefont {Kharzeev}\ and\ \citenamefont
  {Son}(2011)}]{Kharzeev:2010gr}%
  \BibitemOpen
  \bibfield  {author} {\bibinfo {author} {\bibfnamefont {D.~E.}\ \bibnamefont
  {Kharzeev}}\ and\ \bibinfo {author} {\bibfnamefont {D.~T.}\ \bibnamefont
  {Son}},\ }\href {\doibase 10.1103/PhysRevLett.106.062301} {\bibfield
  {journal} {\bibinfo  {journal} {Phys. Rev. Lett.}\ }\textbf {\bibinfo
  {volume} {106}},\ \bibinfo {pages} {062301} (\bibinfo {year} {2011})},\
  \Eprint {http://arxiv.org/abs/1010.0038} {arXiv:1010.0038 [hep-ph]}
  \BibitemShut {NoStop}%
\bibitem [{\citenamefont {Voloshin}(2004)}]{Voloshin:2004vk}%
  \BibitemOpen
  \bibfield  {author} {\bibinfo {author} {\bibfnamefont {S.~A.}\ \bibnamefont
  {Voloshin}},\ }\href {\doibase 10.1103/PhysRevC.70.057901} {\bibfield
  {journal} {\bibinfo  {journal} {Phys. Rev. C}\ }\textbf {\bibinfo {volume}
  {70}},\ \bibinfo {pages} {057901} (\bibinfo {year} {2004})},\ \Eprint
  {http://arxiv.org/abs/hep-ph/0406311} {arXiv:hep-ph/0406311} \BibitemShut
  {NoStop}%
\bibitem [{\citenamefont {Abelev}\ \emph {et~al.}(2009)\citenamefont {Abelev}
  \emph {et~al.}}]{STAR:2009wot}%
  \BibitemOpen
  \bibfield  {author} {\bibinfo {author} {\bibfnamefont {B.~I.}\ \bibnamefont
  {Abelev}} \emph {et~al.} (\bibinfo {collaboration} {STAR}),\ }\href {\doibase
  10.1103/PhysRevLett.103.251601} {\bibfield  {journal} {\bibinfo  {journal}
  {Phys. Rev. Lett.}\ }\textbf {\bibinfo {volume} {103}},\ \bibinfo {pages}
  {251601} (\bibinfo {year} {2009})},\ \Eprint {http://arxiv.org/abs/0909.1739}
  {arXiv:0909.1739 [nucl-ex]} \BibitemShut {NoStop}%
\bibitem [{\citenamefont {Abelev}\ \emph {et~al.}(2010)\citenamefont {Abelev}
  \emph {et~al.}}]{STAR:2009tro}%
  \BibitemOpen
  \bibfield  {author} {\bibinfo {author} {\bibfnamefont {B.~I.}\ \bibnamefont
  {Abelev}} \emph {et~al.} (\bibinfo {collaboration} {STAR}),\ }\href {\doibase
  10.1103/PhysRevC.81.054908} {\bibfield  {journal} {\bibinfo  {journal} {Phys.
  Rev. C}\ }\textbf {\bibinfo {volume} {81}},\ \bibinfo {pages} {054908}
  (\bibinfo {year} {2010})},\ \Eprint {http://arxiv.org/abs/0909.1717}
  {arXiv:0909.1717 [nucl-ex]} \BibitemShut {NoStop}%
\bibitem [{\citenamefont {Aboona}\ \emph {et~al.}(2023)\citenamefont {Aboona}
  \emph {et~al.}}]{STAR:2022ahj}%
  \BibitemOpen
  \bibfield  {author} {\bibinfo {author} {\bibfnamefont {B.}~\bibnamefont
  {Aboona}} \emph {et~al.} (\bibinfo {collaboration} {STAR}),\ }\href {\doibase
  10.1016/j.physletb.2023.137779} {\bibfield  {journal} {\bibinfo  {journal}
  {Phys. Lett. B}\ }\textbf {\bibinfo {volume} {839}},\ \bibinfo {pages}
  {137779} (\bibinfo {year} {2023})},\ \Eprint
  {http://arxiv.org/abs/2209.03467} {arXiv:2209.03467 [nucl-ex]} \BibitemShut
  {NoStop}%
\bibitem [{\citenamefont {Wang}(2010)}]{Wang:2009kd}%
  \BibitemOpen
  \bibfield  {author} {\bibinfo {author} {\bibfnamefont {F.}~\bibnamefont
  {Wang}},\ }\href {\doibase 10.1103/PhysRevC.81.064902} {\bibfield  {journal}
  {\bibinfo  {journal} {Phys. Rev. C}\ }\textbf {\bibinfo {volume} {81}},\
  \bibinfo {pages} {064902} (\bibinfo {year} {2010})},\ \Eprint
  {http://arxiv.org/abs/0911.1482} {arXiv:0911.1482 [nucl-ex]} \BibitemShut
  {NoStop}%
\bibitem [{\citenamefont {Bzdak}\ \emph {et~al.}(2010)\citenamefont {Bzdak},
  \citenamefont {Koch},\ and\ \citenamefont {Liao}}]{Bzdak:2009fc}%
  \BibitemOpen
  \bibfield  {author} {\bibinfo {author} {\bibfnamefont {A.}~\bibnamefont
  {Bzdak}}, \bibinfo {author} {\bibfnamefont {V.}~\bibnamefont {Koch}}, \ and\
  \bibinfo {author} {\bibfnamefont {J.}~\bibnamefont {Liao}},\ }\href {\doibase
  10.1103/PhysRevC.81.031901} {\bibfield  {journal} {\bibinfo  {journal} {Phys.
  Rev. C}\ }\textbf {\bibinfo {volume} {81}},\ \bibinfo {pages} {031901}
  (\bibinfo {year} {2010})},\ \Eprint {http://arxiv.org/abs/0912.5050}
  {arXiv:0912.5050 [nucl-th]} \BibitemShut {NoStop}%
\bibitem [{\citenamefont {Schlichting}\ and\ \citenamefont
  {Pratt}(2011)}]{Schlichting:2010qia}%
  \BibitemOpen
  \bibfield  {author} {\bibinfo {author} {\bibfnamefont {S.}~\bibnamefont
  {Schlichting}}\ and\ \bibinfo {author} {\bibfnamefont {S.}~\bibnamefont
  {Pratt}},\ }\href {\doibase 10.1103/PhysRevC.83.014913} {\bibfield  {journal}
  {\bibinfo  {journal} {Phys. Rev. C}\ }\textbf {\bibinfo {volume} {83}},\
  \bibinfo {pages} {014913} (\bibinfo {year} {2011})},\ \Eprint
  {http://arxiv.org/abs/1009.4283} {arXiv:1009.4283 [nucl-th]} \BibitemShut
  {NoStop}%
\bibitem [{\citenamefont {Pratt}\ \emph {et~al.}(2011)\citenamefont {Pratt},
  \citenamefont {Schlichting},\ and\ \citenamefont {Gavin}}]{Pratt:2010zn}%
  \BibitemOpen
  \bibfield  {author} {\bibinfo {author} {\bibfnamefont {S.}~\bibnamefont
  {Pratt}}, \bibinfo {author} {\bibfnamefont {S.}~\bibnamefont {Schlichting}},
  \ and\ \bibinfo {author} {\bibfnamefont {S.}~\bibnamefont {Gavin}},\ }\href
  {\doibase 10.1103/PhysRevC.84.024909} {\bibfield  {journal} {\bibinfo
  {journal} {Phys. Rev. C}\ }\textbf {\bibinfo {volume} {84}},\ \bibinfo
  {pages} {024909} (\bibinfo {year} {2011})},\ \Eprint
  {http://arxiv.org/abs/1011.6053} {arXiv:1011.6053 [nucl-th]} \BibitemShut
  {NoStop}%
\bibitem [{\citenamefont {Wu}\ \emph {et~al.}(2023)\citenamefont {Wu} \emph
  {et~al.}}]{Wu:2022fwz}%
  \BibitemOpen
  \bibfield  {author} {\bibinfo {author} {\bibfnamefont {W.-Y.}\ \bibnamefont
  {Wu}} \emph {et~al.},\ }\href {\doibase 10.1103/PhysRevC.107.L031902}
  {\bibfield  {journal} {\bibinfo  {journal} {Phys. Rev. C}\ }\textbf {\bibinfo
  {volume} {107}},\ \bibinfo {pages} {L031902} (\bibinfo {year} {2023})},\
  \Eprint {http://arxiv.org/abs/2211.15446} {arXiv:2211.15446 [nucl-th]}
  \BibitemShut {NoStop}%
\bibitem [{\citenamefont {STAR}(2023)}]{STAR:2023ioo}%
  \BibitemOpen
  \bibfield  {author} {\bibinfo {author} {\bibnamefont {STAR}} (\bibinfo
  {collaboration} {STAR}),\ }\href@noop {} {\  (\bibinfo {year} {2023})},\
  \Eprint {http://arxiv.org/abs/2310.13096} {arXiv:2310.13096 [nucl-ex]}
  \BibitemShut {NoStop}%
\bibitem [{\citenamefont {Abdallah}\ \emph {et~al.}(2022)\citenamefont
  {Abdallah} \emph {et~al.}}]{STAR:2021mii}%
  \BibitemOpen
  \bibfield  {author} {\bibinfo {author} {\bibfnamefont {M.}~\bibnamefont
  {Abdallah}} \emph {et~al.} (\bibinfo {collaboration} {STAR}),\ }\href
  {\doibase 10.1103/PhysRevC.105.014901} {\bibfield  {journal} {\bibinfo
  {journal} {Phys. Rev. C}\ }\textbf {\bibinfo {volume} {105}},\ \bibinfo
  {pages} {014901} (\bibinfo {year} {2022})},\ \Eprint
  {http://arxiv.org/abs/2109.00131} {arXiv:2109.00131 [nucl-ex]} \BibitemShut
  {NoStop}%
\bibitem [{\citenamefont {Acharya}\ \emph
  {et~al.}(2020{\natexlab{a}})\citenamefont {Acharya} \emph
  {et~al.}}]{ALICE:2020siw}%
  \BibitemOpen
  \bibfield  {author} {\bibinfo {author} {\bibfnamefont {S.}~\bibnamefont
  {Acharya}} \emph {et~al.} (\bibinfo {collaboration} {ALICE}),\ }\href
  {\doibase 10.1007/JHEP09(2020)160} {\bibfield  {journal} {\bibinfo  {journal}
  {JHEP}\ }\textbf {\bibinfo {volume} {09}},\ \bibinfo {pages} {160} (\bibinfo
  {year} {2020}{\natexlab{a}})},\ \Eprint {http://arxiv.org/abs/2005.14640}
  {arXiv:2005.14640 [nucl-ex]} \BibitemShut {NoStop}%
\bibitem [{\citenamefont {Xu}(2023)}]{Xu:2023wcy}%
  \BibitemOpen
  \bibfield  {author} {\bibinfo {author} {\bibfnamefont {Z.}~\bibnamefont {Xu}}
  (\bibinfo {collaboration} {STAR}),\ }in\ \href@noop {} {\emph {\bibinfo
  {booktitle} {{30th International Conference on Ultrarelativstic
  Nucleus-Nucleus Collisions}}}}\ (\bibinfo {year} {2023})\ \Eprint
  {http://arxiv.org/abs/2401.00317} {arXiv:2401.00317 [nucl-ex]} \BibitemShut
  {NoStop}%
\bibitem [{\citenamefont {Xu}\ \emph {et~al.}(2024)\citenamefont {Xu},
  \citenamefont {Chan}, \citenamefont {Wang}, \citenamefont {Tang},\ and\
  \citenamefont {Huang}}]{Xu:2023elq}%
  \BibitemOpen
  \bibfield  {author} {\bibinfo {author} {\bibfnamefont {Z.}~\bibnamefont
  {Xu}}, \bibinfo {author} {\bibfnamefont {B.}~\bibnamefont {Chan}}, \bibinfo
  {author} {\bibfnamefont {G.}~\bibnamefont {Wang}}, \bibinfo {author}
  {\bibfnamefont {A.}~\bibnamefont {Tang}}, \ and\ \bibinfo {author}
  {\bibfnamefont {H.~Z.}\ \bibnamefont {Huang}},\ }\href {\doibase
  10.1016/j.physletb.2023.138367} {\bibfield  {journal} {\bibinfo  {journal}
  {Phys. Lett. B}\ }\textbf {\bibinfo {volume} {848}},\ \bibinfo {pages}
  {138367} (\bibinfo {year} {2024})},\ \Eprint
  {http://arxiv.org/abs/2307.14997} {arXiv:2307.14997 [nucl-th]} \BibitemShut
  {NoStop}%
\bibitem [{\citenamefont {Wang}(2023)}]{Wang:2023xhn}%
  \BibitemOpen
  \bibfield  {author} {\bibinfo {author} {\bibfnamefont {C.-Z.}\ \bibnamefont
  {Wang}} (\bibinfo {collaboration} {ALICE}),\ }\href@noop {} {\  (\bibinfo
  {year} {2023})},\ \Eprint {http://arxiv.org/abs/2312.07346} {arXiv:2312.07346
  [nucl-ex]} \BibitemShut {NoStop}%
\bibitem [{\citenamefont {Yuan}\ \emph {et~al.}(2023)\citenamefont {Yuan},
  \citenamefont {Huang}, \citenamefont {Zhou}, \citenamefont {Ma},\ and\
  \citenamefont {Huang}}]{Yuan:2023skl}%
  \BibitemOpen
  \bibfield  {author} {\bibinfo {author} {\bibfnamefont {Z.}~\bibnamefont
  {Yuan}}, \bibinfo {author} {\bibfnamefont {A.}~\bibnamefont {Huang}},
  \bibinfo {author} {\bibfnamefont {W.-H.}\ \bibnamefont {Zhou}}, \bibinfo
  {author} {\bibfnamefont {G.-L.}\ \bibnamefont {Ma}}, \ and\ \bibinfo {author}
  {\bibfnamefont {M.}~\bibnamefont {Huang}},\ }\href@noop {} {\  (\bibinfo
  {year} {2023})},\ \Eprint {http://arxiv.org/abs/2310.20194} {arXiv:2310.20194
  [hep-ph]} \BibitemShut {NoStop}%
\bibitem [{\citenamefont {Yu}\ \emph {et~al.}(2014)\citenamefont {Yu},
  \citenamefont {Liu},\ and\ \citenamefont {Huang}}]{Yu:2014sla}%
  \BibitemOpen
  \bibfield  {author} {\bibinfo {author} {\bibfnamefont {L.}~\bibnamefont
  {Yu}}, \bibinfo {author} {\bibfnamefont {H.}~\bibnamefont {Liu}}, \ and\
  \bibinfo {author} {\bibfnamefont {M.}~\bibnamefont {Huang}},\ }\href
  {\doibase 10.1103/PhysRevD.90.074009} {\bibfield  {journal} {\bibinfo
  {journal} {Phys. Rev. D}\ }\textbf {\bibinfo {volume} {90}},\ \bibinfo
  {pages} {074009} (\bibinfo {year} {2014})},\ \Eprint
  {http://arxiv.org/abs/1404.6969} {arXiv:1404.6969 [hep-ph]} \BibitemShut
  {NoStop}%
\bibitem [{\citenamefont {Bjorken}(1983)}]{Bjorken:1982qr}%
  \BibitemOpen
  \bibfield  {author} {\bibinfo {author} {\bibfnamefont {J.~D.}\ \bibnamefont
  {Bjorken}},\ }\href {\doibase 10.1103/PhysRevD.27.140} {\bibfield  {journal}
  {\bibinfo  {journal} {Phys. Rev. D}\ }\textbf {\bibinfo {volume} {27}},\
  \bibinfo {pages} {140} (\bibinfo {year} {1983})}\BibitemShut {NoStop}%
\bibitem [{\citenamefont {Florkowski}(2010)}]{Florkowski:2010zz}%
  \BibitemOpen
  \bibfield  {author} {\bibinfo {author} {\bibfnamefont {W.}~\bibnamefont
  {Florkowski}},\ }\href@noop {} {\emph {\bibinfo {title} {{Phenomenology of
  Ultra-Relativistic Heavy-Ion Collisions}}}}\ (\bibinfo {year}
  {2010})\BibitemShut {NoStop}%
\bibitem [{\citenamefont {Floerchinger}\ and\ \citenamefont
  {Martinez}(2015)}]{Floerchinger:2015efa}%
  \BibitemOpen
  \bibfield  {author} {\bibinfo {author} {\bibfnamefont {S.}~\bibnamefont
  {Floerchinger}}\ and\ \bibinfo {author} {\bibfnamefont {M.}~\bibnamefont
  {Martinez}},\ }\href {\doibase 10.1103/PhysRevC.92.064906} {\bibfield
  {journal} {\bibinfo  {journal} {Phys. Rev. C}\ }\textbf {\bibinfo {volume}
  {92}},\ \bibinfo {pages} {064906} (\bibinfo {year} {2015})},\ \Eprint
  {http://arxiv.org/abs/1507.05569} {arXiv:1507.05569 [nucl-th]} \BibitemShut
  {NoStop}%
\bibitem [{\citenamefont {Jiang}\ \emph {et~al.}(2016)\citenamefont {Jiang},
  \citenamefont {Lin},\ and\ \citenamefont {Liao}}]{Jiang:2016woz}%
  \BibitemOpen
  \bibfield  {author} {\bibinfo {author} {\bibfnamefont {Y.}~\bibnamefont
  {Jiang}}, \bibinfo {author} {\bibfnamefont {Z.-W.}\ \bibnamefont {Lin}}, \
  and\ \bibinfo {author} {\bibfnamefont {J.}~\bibnamefont {Liao}},\ }\href
  {\doibase 10.1103/PhysRevC.94.044910} {\bibfield  {journal} {\bibinfo
  {journal} {Phys. Rev. C}\ }\textbf {\bibinfo {volume} {94}},\ \bibinfo
  {pages} {044910} (\bibinfo {year} {2016})},\ \bibinfo {note} {[Erratum:
  Phys.Rev.C 95, 049904 (2017)]},\ \Eprint {http://arxiv.org/abs/1602.06580}
  {arXiv:1602.06580 [hep-ph]} \BibitemShut {NoStop}%
\bibitem [{\citenamefont {Deng}\ and\ \citenamefont
  {Huang}(2016)}]{Deng:2016gyh}%
  \BibitemOpen
  \bibfield  {author} {\bibinfo {author} {\bibfnamefont {W.-T.}\ \bibnamefont
  {Deng}}\ and\ \bibinfo {author} {\bibfnamefont {X.-G.}\ \bibnamefont
  {Huang}},\ }\href {\doibase 10.1103/PhysRevC.93.064907} {\bibfield  {journal}
  {\bibinfo  {journal} {Phys. Rev. C}\ }\textbf {\bibinfo {volume} {93}},\
  \bibinfo {pages} {064907} (\bibinfo {year} {2016})},\ \Eprint
  {http://arxiv.org/abs/1603.06117} {arXiv:1603.06117 [nucl-th]} \BibitemShut
  {NoStop}%
\bibitem [{\citenamefont {Karpenko}(2021)}]{Karpenko:2021wdm}%
  \BibitemOpen
  \bibfield  {author} {\bibinfo {author} {\bibfnamefont {I.}~\bibnamefont
  {Karpenko}},\ }\enquote {\bibinfo {title} {{Vorticity and Polarization in
  Heavy-Ion Collisions: Hydrodynamic Models}},}\ \ (\bibinfo {year} {2021})\
  \Eprint {http://arxiv.org/abs/2101.04963} {arXiv:2101.04963 [nucl-th]}
  \BibitemShut {NoStop}%
\bibitem [{\citenamefont {Huang}\ \emph {et~al.}(2021)\citenamefont {Huang},
  \citenamefont {Liao}, \citenamefont {Wang},\ and\ \citenamefont
  {Xia}}]{Huang:2020dtn}%
  \BibitemOpen
  \bibfield  {author} {\bibinfo {author} {\bibfnamefont {X.-G.}\ \bibnamefont
  {Huang}}, \bibinfo {author} {\bibfnamefont {J.}~\bibnamefont {Liao}},
  \bibinfo {author} {\bibfnamefont {Q.}~\bibnamefont {Wang}}, \ and\ \bibinfo
  {author} {\bibfnamefont {X.-L.}\ \bibnamefont {Xia}},\ }\href {\doibase
  10.1007/978-3-030-71427-7_9} {\bibfield  {journal} {\bibinfo  {journal}
  {Lect. Notes Phys.}\ }\textbf {\bibinfo {volume} {987}},\ \bibinfo {pages}
  {281} (\bibinfo {year} {2021})},\ \Eprint {http://arxiv.org/abs/2010.08937}
  {arXiv:2010.08937 [nucl-th]} \BibitemShut {NoStop}%
\bibitem [{\citenamefont {Ivanov}\ and\ \citenamefont
  {Soldatov}(2017)}]{Ivanov:2017dff}%
  \BibitemOpen
  \bibfield  {author} {\bibinfo {author} {\bibfnamefont {Y.~B.}\ \bibnamefont
  {Ivanov}}\ and\ \bibinfo {author} {\bibfnamefont {A.~A.}\ \bibnamefont
  {Soldatov}},\ }\href {\doibase 10.1103/PhysRevC.95.054915} {\bibfield
  {journal} {\bibinfo  {journal} {Phys. Rev. C}\ }\textbf {\bibinfo {volume}
  {95}},\ \bibinfo {pages} {054915} (\bibinfo {year} {2017})},\ \Eprint
  {http://arxiv.org/abs/1701.01319} {arXiv:1701.01319 [nucl-th]} \BibitemShut
  {NoStop}%
\bibitem [{\citenamefont {Deng}\ \emph {et~al.}(2020)\citenamefont {Deng},
  \citenamefont {Huang}, \citenamefont {Ma},\ and\ \citenamefont
  {Zhang}}]{Deng:2020ygd}%
  \BibitemOpen
  \bibfield  {author} {\bibinfo {author} {\bibfnamefont {X.-G.}\ \bibnamefont
  {Deng}}, \bibinfo {author} {\bibfnamefont {X.-G.}\ \bibnamefont {Huang}},
  \bibinfo {author} {\bibfnamefont {Y.-G.}\ \bibnamefont {Ma}}, \ and\ \bibinfo
  {author} {\bibfnamefont {S.}~\bibnamefont {Zhang}},\ }\href {\doibase
  10.1103/PhysRevC.101.064908} {\bibfield  {journal} {\bibinfo  {journal}
  {Phys. Rev. C}\ }\textbf {\bibinfo {volume} {101}},\ \bibinfo {pages}
  {064908} (\bibinfo {year} {2020})},\ \Eprint
  {http://arxiv.org/abs/2001.01371} {arXiv:2001.01371 [nucl-th]} \BibitemShut
  {NoStop}%
\bibitem [{\citenamefont {Lin}\ \emph {et~al.}(2005)\citenamefont {Lin},
  \citenamefont {Ko}, \citenamefont {Li}, \citenamefont {Zhang},\ and\
  \citenamefont {Pal}}]{Lin:2004en}%
  \BibitemOpen
  \bibfield  {author} {\bibinfo {author} {\bibfnamefont {Z.-W.}\ \bibnamefont
  {Lin}}, \bibinfo {author} {\bibfnamefont {C.~M.}\ \bibnamefont {Ko}},
  \bibinfo {author} {\bibfnamefont {B.-A.}\ \bibnamefont {Li}}, \bibinfo
  {author} {\bibfnamefont {B.}~\bibnamefont {Zhang}}, \ and\ \bibinfo {author}
  {\bibfnamefont {S.}~\bibnamefont {Pal}},\ }\href {\doibase
  10.1103/PhysRevC.72.064901} {\bibfield  {journal} {\bibinfo  {journal} {Phys.
  Rev. C}\ }\textbf {\bibinfo {volume} {72}},\ \bibinfo {pages} {064901}
  (\bibinfo {year} {2005})},\ \Eprint {http://arxiv.org/abs/nucl-th/0411110}
  {arXiv:nucl-th/0411110} \BibitemShut {NoStop}%
\bibitem [{\citenamefont {Adamczyk}\ \emph {et~al.}(2013)\citenamefont
  {Adamczyk} \emph {et~al.}}]{STAR:2013ksd}%
  \BibitemOpen
  \bibfield  {author} {\bibinfo {author} {\bibfnamefont {L.}~\bibnamefont
  {Adamczyk}} \emph {et~al.} (\bibinfo {collaboration} {STAR}),\ }\href
  {\doibase 10.1103/PhysRevC.88.064911} {\bibfield  {journal} {\bibinfo
  {journal} {Phys. Rev. C}\ }\textbf {\bibinfo {volume} {88}},\ \bibinfo
  {pages} {064911} (\bibinfo {year} {2013})},\ \Eprint
  {http://arxiv.org/abs/1302.3802} {arXiv:1302.3802 [nucl-ex]} \BibitemShut
  {NoStop}%
\bibitem [{\citenamefont {Adamczyk}\ \emph {et~al.}(2014)\citenamefont
  {Adamczyk} \emph {et~al.}}]{STAR:2014uiw}%
  \BibitemOpen
  \bibfield  {author} {\bibinfo {author} {\bibfnamefont {L.}~\bibnamefont
  {Adamczyk}} \emph {et~al.} (\bibinfo {collaboration} {STAR}),\ }\href
  {\doibase 10.1103/PhysRevLett.113.052302} {\bibfield  {journal} {\bibinfo
  {journal} {Phys. Rev. Lett.}\ }\textbf {\bibinfo {volume} {113}},\ \bibinfo
  {pages} {052302} (\bibinfo {year} {2014})},\ \Eprint
  {http://arxiv.org/abs/1404.1433} {arXiv:1404.1433 [nucl-ex]} \BibitemShut
  {NoStop}%
\bibitem [{\citenamefont {Sirunyan}\ \emph {et~al.}(2018)\citenamefont
  {Sirunyan} \emph {et~al.}}]{CMS:2017lrw}%
  \BibitemOpen
  \bibfield  {author} {\bibinfo {author} {\bibfnamefont {A.~M.}\ \bibnamefont
  {Sirunyan}} \emph {et~al.} (\bibinfo {collaboration} {CMS}),\ }\href
  {\doibase 10.1103/PhysRevC.97.044912} {\bibfield  {journal} {\bibinfo
  {journal} {Phys. Rev. C}\ }\textbf {\bibinfo {volume} {97}},\ \bibinfo
  {pages} {044912} (\bibinfo {year} {2018})},\ \Eprint
  {http://arxiv.org/abs/1708.01602} {arXiv:1708.01602 [nucl-ex]} \BibitemShut
  {NoStop}%
\bibitem [{\citenamefont {ALICE}(2022)}]{ALICE:2022ljz}%
  \BibitemOpen
  \bibfield  {author} {\bibinfo {author} {\bibnamefont {ALICE}} (\bibinfo
  {collaboration} {ALICE}),\ }\href@noop {} {\  (\bibinfo {year} {2022})},\
  \Eprint {http://arxiv.org/abs/2210.15383} {arXiv:2210.15383 [nucl-ex]}
  \BibitemShut {NoStop}%
\bibitem [{\citenamefont {Sch\"afer}\ \emph {et~al.}(2022)\citenamefont
  {Sch\"afer}, \citenamefont {Karpenko}, \citenamefont {Wu}, \citenamefont
  {Hammelmann},\ and\ \citenamefont {Elfner}}]{Schafer:2021csj}%
  \BibitemOpen
  \bibfield  {author} {\bibinfo {author} {\bibfnamefont {A.}~\bibnamefont
  {Sch\"afer}}, \bibinfo {author} {\bibfnamefont {I.}~\bibnamefont {Karpenko}},
  \bibinfo {author} {\bibfnamefont {X.-Y.}\ \bibnamefont {Wu}}, \bibinfo
  {author} {\bibfnamefont {J.}~\bibnamefont {Hammelmann}}, \ and\ \bibinfo
  {author} {\bibfnamefont {H.}~\bibnamefont {Elfner}} (\bibinfo {collaboration}
  {SMASH}),\ }\href {\doibase 10.1140/epja/s10050-022-00872-x} {\bibfield
  {journal} {\bibinfo  {journal} {Eur. Phys. J. A}\ }\textbf {\bibinfo {volume}
  {58}},\ \bibinfo {pages} {230} (\bibinfo {year} {2022})},\ \Eprint
  {http://arxiv.org/abs/2112.08724} {arXiv:2112.08724 [hep-ph]} \BibitemShut
  {NoStop}%
\bibitem [{\citenamefont {Ciacco}(2023)}]{Ciacco:2023ekv}%
  \BibitemOpen
  \bibfield  {author} {\bibinfo {author} {\bibfnamefont {M.}~\bibnamefont
  {Ciacco}} (\bibinfo {collaboration} {ALICE})\ }(\bibinfo {year} {2023})\
  \Eprint {http://arxiv.org/abs/2301.11091} {arXiv:2301.11091 [nucl-ex]}
  \BibitemShut {NoStop}%
\bibitem [{\citenamefont {Acharya}\ \emph {et~al.}(2023)\citenamefont {Acharya}
  \emph {et~al.}}]{ALICE:2023ulv}%
  \BibitemOpen
  \bibfield  {author} {\bibinfo {author} {\bibfnamefont {S.}~\bibnamefont
  {Acharya}} \emph {et~al.} (\bibinfo {collaboration} {ALICE}),\ }\href@noop {}
  {\  (\bibinfo {year} {2023})},\ \Eprint {http://arxiv.org/abs/2311.13332}
  {arXiv:2311.13332 [nucl-ex]} \BibitemShut {NoStop}%
\bibitem [{\citenamefont {Zakharov}(2021)}]{Zakharov:2021uza}%
  \BibitemOpen
  \bibfield  {author} {\bibinfo {author} {\bibfnamefont {B.~G.}\ \bibnamefont
  {Zakharov}},\ }\href {\doibase 10.1007/JHEP09(2021)087} {\bibfield  {journal}
  {\bibinfo  {journal} {JHEP}\ }\textbf {\bibinfo {volume} {09}},\ \bibinfo
  {pages} {087} (\bibinfo {year} {2021})},\ \Eprint
  {http://arxiv.org/abs/2105.09350} {arXiv:2105.09350 [hep-ph]} \BibitemShut
  {NoStop}%
\bibitem [{\citenamefont {Loizides}\ \emph {et~al.}(2018)\citenamefont
  {Loizides}, \citenamefont {Kamin},\ and\ \citenamefont
  {d'Enterria}}]{Loizides:2017ack}%
  \BibitemOpen
  \bibfield  {author} {\bibinfo {author} {\bibfnamefont {C.}~\bibnamefont
  {Loizides}}, \bibinfo {author} {\bibfnamefont {J.}~\bibnamefont {Kamin}}, \
  and\ \bibinfo {author} {\bibfnamefont {D.}~\bibnamefont {d'Enterria}},\
  }\href {\doibase 10.1103/PhysRevC.97.054910} {\bibfield  {journal} {\bibinfo
  {journal} {Phys. Rev. C}\ }\textbf {\bibinfo {volume} {97}},\ \bibinfo
  {pages} {054910} (\bibinfo {year} {2018})},\ \bibinfo {note} {[Erratum:
  Phys.Rev.C 99, 019901 (2019)]},\ \Eprint {http://arxiv.org/abs/1710.07098}
  {arXiv:1710.07098 [nucl-ex]} \BibitemShut {NoStop}%
\bibitem [{\citenamefont {Acharya}\ \emph
  {et~al.}(2020{\natexlab{b}})\citenamefont {Acharya} \emph
  {et~al.}}]{ALICE:2019hno}%
  \BibitemOpen
  \bibfield  {author} {\bibinfo {author} {\bibfnamefont {S.}~\bibnamefont
  {Acharya}} \emph {et~al.} (\bibinfo {collaboration} {ALICE}),\ }\href
  {\doibase 10.1103/PhysRevC.101.044907} {\bibfield  {journal} {\bibinfo
  {journal} {Phys. Rev. C}\ }\textbf {\bibinfo {volume} {101}},\ \bibinfo
  {pages} {044907} (\bibinfo {year} {2020}{\natexlab{b}})},\ \Eprint
  {http://arxiv.org/abs/1910.07678} {arXiv:1910.07678 [nucl-ex]} \BibitemShut
  {NoStop}%
\bibitem [{\citenamefont {McLerran}\ and\ \citenamefont
  {Skokov}(2014)}]{McLerran:2013hla}%
  \BibitemOpen
  \bibfield  {author} {\bibinfo {author} {\bibfnamefont {L.}~\bibnamefont
  {McLerran}}\ and\ \bibinfo {author} {\bibfnamefont {V.}~\bibnamefont
  {Skokov}},\ }\href {\doibase 10.1016/j.nuclphysa.2014.05.008} {\bibfield
  {journal} {\bibinfo  {journal} {Nucl. Phys. A}\ }\textbf {\bibinfo {volume}
  {929}},\ \bibinfo {pages} {184} (\bibinfo {year} {2014})},\ \Eprint
  {http://arxiv.org/abs/1305.0774} {arXiv:1305.0774 [hep-ph]} \BibitemShut
  {NoStop}%
\bibitem [{\citenamefont {G\"ursoy}\ \emph {et~al.}(2018)\citenamefont
  {G\"ursoy}, \citenamefont {Kharzeev}, \citenamefont {Marcus}, \citenamefont
  {Rajagopal},\ and\ \citenamefont {Shen}}]{Gursoy:2018yai}%
  \BibitemOpen
  \bibfield  {author} {\bibinfo {author} {\bibfnamefont {U.}~\bibnamefont
  {G\"ursoy}}, \bibinfo {author} {\bibfnamefont {D.}~\bibnamefont {Kharzeev}},
  \bibinfo {author} {\bibfnamefont {E.}~\bibnamefont {Marcus}}, \bibinfo
  {author} {\bibfnamefont {K.}~\bibnamefont {Rajagopal}}, \ and\ \bibinfo
  {author} {\bibfnamefont {C.}~\bibnamefont {Shen}},\ }\href {\doibase
  10.1103/PhysRevC.98.055201} {\bibfield  {journal} {\bibinfo  {journal} {Phys.
  Rev. C}\ }\textbf {\bibinfo {volume} {98}},\ \bibinfo {pages} {055201}
  (\bibinfo {year} {2018})},\ \Eprint {http://arxiv.org/abs/1806.05288}
  {arXiv:1806.05288 [hep-ph]} \BibitemShut {NoStop}%
\bibitem [{\citenamefont {Hattori}\ and\ \citenamefont
  {Huang}(2017)}]{Hattori:2016emy}%
  \BibitemOpen
  \bibfield  {author} {\bibinfo {author} {\bibfnamefont {K.}~\bibnamefont
  {Hattori}}\ and\ \bibinfo {author} {\bibfnamefont {X.-G.}\ \bibnamefont
  {Huang}},\ }\href {\doibase 10.1007/s41365-016-0178-3} {\bibfield  {journal}
  {\bibinfo  {journal} {Nucl. Sci. Tech.}\ }\textbf {\bibinfo {volume} {28}},\
  \bibinfo {pages} {26} (\bibinfo {year} {2017})},\ \Eprint
  {http://arxiv.org/abs/1609.00747} {arXiv:1609.00747 [nucl-th]} \BibitemShut
  {NoStop}%
\bibitem [{\citenamefont {Huang}(2016)}]{Huang:2015oca}%
  \BibitemOpen
  \bibfield  {author} {\bibinfo {author} {\bibfnamefont {X.-G.}\ \bibnamefont
  {Huang}},\ }\href {\doibase 10.1088/0034-4885/79/7/076302} {\bibfield
  {journal} {\bibinfo  {journal} {Rept. Prog. Phys.}\ }\textbf {\bibinfo
  {volume} {79}},\ \bibinfo {pages} {076302} (\bibinfo {year} {2016})},\
  \Eprint {http://arxiv.org/abs/1509.04073} {arXiv:1509.04073 [nucl-th]}
  \BibitemShut {NoStop}%
\bibitem [{\citenamefont {Dash}\ \emph {et~al.}(2023)\citenamefont {Dash},
  \citenamefont {Shokri}, \citenamefont {Rezzolla},\ and\ \citenamefont
  {Rischke}}]{Dash:2022xkz}%
  \BibitemOpen
  \bibfield  {author} {\bibinfo {author} {\bibfnamefont {A.}~\bibnamefont
  {Dash}}, \bibinfo {author} {\bibfnamefont {M.}~\bibnamefont {Shokri}},
  \bibinfo {author} {\bibfnamefont {L.}~\bibnamefont {Rezzolla}}, \ and\
  \bibinfo {author} {\bibfnamefont {D.~H.}\ \bibnamefont {Rischke}},\ }\href
  {\doibase 10.1103/PhysRevD.107.056003} {\bibfield  {journal} {\bibinfo
  {journal} {Phys. Rev. D}\ }\textbf {\bibinfo {volume} {107}},\ \bibinfo
  {pages} {056003} (\bibinfo {year} {2023})},\ \Eprint
  {http://arxiv.org/abs/2211.09459} {arXiv:2211.09459 [nucl-th]} \BibitemShut
  {NoStop}%
\bibitem [{\citenamefont {Shen}\ \emph {et~al.}(2016)\citenamefont {Shen},
  \citenamefont {Qiu}, \citenamefont {Song}, \citenamefont {Bernhard},
  \citenamefont {Bass},\ and\ \citenamefont {Heinz}}]{Shen:2014vra}%
  \BibitemOpen
  \bibfield  {author} {\bibinfo {author} {\bibfnamefont {C.}~\bibnamefont
  {Shen}}, \bibinfo {author} {\bibfnamefont {Z.}~\bibnamefont {Qiu}}, \bibinfo
  {author} {\bibfnamefont {H.}~\bibnamefont {Song}}, \bibinfo {author}
  {\bibfnamefont {J.}~\bibnamefont {Bernhard}}, \bibinfo {author}
  {\bibfnamefont {S.}~\bibnamefont {Bass}}, \ and\ \bibinfo {author}
  {\bibfnamefont {U.}~\bibnamefont {Heinz}},\ }\href {\doibase
  10.1016/j.cpc.2015.08.039} {\bibfield  {journal} {\bibinfo  {journal}
  {Comput. Phys. Commun.}\ }\textbf {\bibinfo {volume} {199}},\ \bibinfo
  {pages} {61} (\bibinfo {year} {2016})},\ \Eprint
  {http://arxiv.org/abs/1409.8164} {arXiv:1409.8164 [nucl-th]} \BibitemShut
  {NoStop}%
\bibitem [{\citenamefont {Kaczmarek}\ \emph {et~al.}(2012)\citenamefont
  {Kaczmarek}, \citenamefont {Laermann}, \citenamefont {M\"uller},
  \citenamefont {Karsch}, \citenamefont {Ding}, \citenamefont {Mukherjee},
  \citenamefont {Francis},\ and\ \citenamefont {Soeldner}}]{Kaczmarek:2012mgs}%
  \BibitemOpen
  \bibfield  {author} {\bibinfo {author} {\bibfnamefont {O.}~\bibnamefont
  {Kaczmarek}}, \bibinfo {author} {\bibfnamefont {E.}~\bibnamefont {Laermann}},
  \bibinfo {author} {\bibfnamefont {M.}~\bibnamefont {M\"uller}}, \bibinfo
  {author} {\bibfnamefont {F.}~\bibnamefont {Karsch}}, \bibinfo {author}
  {\bibfnamefont {H.~T.}\ \bibnamefont {Ding}}, \bibinfo {author}
  {\bibfnamefont {S.}~\bibnamefont {Mukherjee}}, \bibinfo {author}
  {\bibfnamefont {A.}~\bibnamefont {Francis}}, \ and\ \bibinfo {author}
  {\bibfnamefont {W.}~\bibnamefont {Soeldner}},\ }\href {\doibase
  10.22323/1.171.0185} {\bibfield  {journal} {\bibinfo  {journal} {PoS}\
  }\textbf {\bibinfo {volume} {ConfinementX}},\ \bibinfo {pages} {185}
  (\bibinfo {year} {2012})},\ \Eprint {http://arxiv.org/abs/1301.7436}
  {arXiv:1301.7436 [hep-lat]} \BibitemShut {NoStop}%
\bibitem [{\citenamefont {Aarts}\ \emph {et~al.}(2007)\citenamefont {Aarts},
  \citenamefont {Allton}, \citenamefont {Foley}, \citenamefont {Hands},\ and\
  \citenamefont {Kim}}]{Aarts:2007wj}%
  \BibitemOpen
  \bibfield  {author} {\bibinfo {author} {\bibfnamefont {G.}~\bibnamefont
  {Aarts}}, \bibinfo {author} {\bibfnamefont {C.}~\bibnamefont {Allton}},
  \bibinfo {author} {\bibfnamefont {J.}~\bibnamefont {Foley}}, \bibinfo
  {author} {\bibfnamefont {S.}~\bibnamefont {Hands}}, \ and\ \bibinfo {author}
  {\bibfnamefont {S.}~\bibnamefont {Kim}},\ }\href {\doibase
  10.1103/PhysRevLett.99.022002} {\bibfield  {journal} {\bibinfo  {journal}
  {Phys. Rev. Lett.}\ }\textbf {\bibinfo {volume} {99}},\ \bibinfo {pages}
  {022002} (\bibinfo {year} {2007})},\ \Eprint
  {http://arxiv.org/abs/hep-lat/0703008} {arXiv:hep-lat/0703008} \BibitemShut
  {NoStop}%
\bibitem [{\citenamefont {Kharzeev}\ and\ \citenamefont
  {Yee}(2011)}]{Kharzeev:2010gd}%
  \BibitemOpen
  \bibfield  {author} {\bibinfo {author} {\bibfnamefont {D.~E.}\ \bibnamefont
  {Kharzeev}}\ and\ \bibinfo {author} {\bibfnamefont {H.-U.}\ \bibnamefont
  {Yee}},\ }\href {\doibase 10.1103/PhysRevD.83.085007} {\bibfield  {journal}
  {\bibinfo  {journal} {Phys. Rev. D}\ }\textbf {\bibinfo {volume} {83}},\
  \bibinfo {pages} {085007} (\bibinfo {year} {2011})},\ \Eprint
  {http://arxiv.org/abs/1012.6026} {arXiv:1012.6026 [hep-th]} \BibitemShut
  {NoStop}%
\bibitem [{\citenamefont {Burnier}\ \emph {et~al.}(2011)\citenamefont
  {Burnier}, \citenamefont {Kharzeev}, \citenamefont {Liao},\ and\
  \citenamefont {Yee}}]{Burnier:2011bf}%
  \BibitemOpen
  \bibfield  {author} {\bibinfo {author} {\bibfnamefont {Y.}~\bibnamefont
  {Burnier}}, \bibinfo {author} {\bibfnamefont {D.~E.}\ \bibnamefont
  {Kharzeev}}, \bibinfo {author} {\bibfnamefont {J.}~\bibnamefont {Liao}}, \
  and\ \bibinfo {author} {\bibfnamefont {H.-U.}\ \bibnamefont {Yee}},\ }\href
  {\doibase 10.1103/PhysRevLett.107.052303} {\bibfield  {journal} {\bibinfo
  {journal} {Phys. Rev. Lett.}\ }\textbf {\bibinfo {volume} {107}},\ \bibinfo
  {pages} {052303} (\bibinfo {year} {2011})},\ \Eprint
  {http://arxiv.org/abs/1103.1307} {arXiv:1103.1307 [hep-ph]} \BibitemShut
  {NoStop}%
\bibitem [{\citenamefont {Wang}(2013)}]{Wang:2012qs}%
  \BibitemOpen
  \bibfield  {author} {\bibinfo {author} {\bibfnamefont {G.}~\bibnamefont
  {Wang}} (\bibinfo {collaboration} {STAR}),\ }\href {\doibase
  10.1016/j.nuclphysa.2013.01.069} {\bibfield  {journal} {\bibinfo  {journal}
  {Nucl. Phys. A}\ }\textbf {\bibinfo {volume} {904-905}},\ \bibinfo {pages}
  {248c} (\bibinfo {year} {2013})},\ \Eprint {http://arxiv.org/abs/1210.5498}
  {arXiv:1210.5498 [nucl-ex]} \BibitemShut {NoStop}%
\bibitem [{\citenamefont {Adamczyk}\ \emph {et~al.}(2015)\citenamefont
  {Adamczyk} \emph {et~al.}}]{STAR:2015wza}%
  \BibitemOpen
  \bibfield  {author} {\bibinfo {author} {\bibfnamefont {L.}~\bibnamefont
  {Adamczyk}} \emph {et~al.} (\bibinfo {collaboration} {STAR}),\ }\href
  {\doibase 10.1103/PhysRevLett.114.252302} {\bibfield  {journal} {\bibinfo
  {journal} {Phys. Rev. Lett.}\ }\textbf {\bibinfo {volume} {114}},\ \bibinfo
  {pages} {252302} (\bibinfo {year} {2015})},\ \Eprint
  {http://arxiv.org/abs/1504.02175} {arXiv:1504.02175 [nucl-ex]} \BibitemShut
  {NoStop}%
\bibitem [{\citenamefont {Kharzeev}(1996)}]{Kharzeev:1996sq}%
  \BibitemOpen
  \bibfield  {author} {\bibinfo {author} {\bibfnamefont {D.}~\bibnamefont
  {Kharzeev}},\ }\href {\doibase 10.1016/0370-2693(96)00435-2} {\bibfield
  {journal} {\bibinfo  {journal} {Phys. Lett. B}\ }\textbf {\bibinfo {volume}
  {378}},\ \bibinfo {pages} {238} (\bibinfo {year} {1996})},\ \Eprint
  {http://arxiv.org/abs/nucl-th/9602027} {arXiv:nucl-th/9602027} \BibitemShut
  {NoStop}%
\bibitem [{\citenamefont {Brandenburg}\ \emph {et~al.}(2022)\citenamefont
  {Brandenburg}, \citenamefont {Lewis}, \citenamefont {Tribedy},\ and\
  \citenamefont {Xu}}]{Brandenburg:2022hrp}%
  \BibitemOpen
  \bibfield  {author} {\bibinfo {author} {\bibfnamefont {J.~D.}\ \bibnamefont
  {Brandenburg}}, \bibinfo {author} {\bibfnamefont {N.}~\bibnamefont {Lewis}},
  \bibinfo {author} {\bibfnamefont {P.}~\bibnamefont {Tribedy}}, \ and\
  \bibinfo {author} {\bibfnamefont {Z.}~\bibnamefont {Xu}},\ }\href@noop {} {\
  (\bibinfo {year} {2022})},\ \Eprint {http://arxiv.org/abs/2205.05685}
  {arXiv:2205.05685 [hep-ph]} \BibitemShut {NoStop}%
\bibitem [{\citenamefont {Lv}\ \emph {et~al.}(2023)\citenamefont {Lv},
  \citenamefont {Li}, \citenamefont {Li}, \citenamefont {Ma}, \citenamefont
  {Tang}, \citenamefont {Tribedy}, \citenamefont {Tsang}, \citenamefont {Xu},\
  and\ \citenamefont {Zha}}]{Lv:2023fxk}%
  \BibitemOpen
  \bibfield  {author} {\bibinfo {author} {\bibfnamefont {W.}~\bibnamefont
  {Lv}}, \bibinfo {author} {\bibfnamefont {Y.}~\bibnamefont {Li}}, \bibinfo
  {author} {\bibfnamefont {Z.}~\bibnamefont {Li}}, \bibinfo {author}
  {\bibfnamefont {R.}~\bibnamefont {Ma}}, \bibinfo {author} {\bibfnamefont
  {Z.}~\bibnamefont {Tang}}, \bibinfo {author} {\bibfnamefont {P.}~\bibnamefont
  {Tribedy}}, \bibinfo {author} {\bibfnamefont {C.~Y.}\ \bibnamefont {Tsang}},
  \bibinfo {author} {\bibfnamefont {Z.}~\bibnamefont {Xu}}, \ and\ \bibinfo
  {author} {\bibfnamefont {W.}~\bibnamefont {Zha}},\ }\href@noop {} {\
  (\bibinfo {year} {2023})},\ \Eprint {http://arxiv.org/abs/2309.06445}
  {arXiv:2309.06445 [nucl-th]} \BibitemShut {NoStop}%
\bibitem [{\citenamefont {Abbas}\ \emph {et~al.}(2013)\citenamefont {Abbas}
  \emph {et~al.}}]{ALICE:2013yba}%
  \BibitemOpen
  \bibfield  {author} {\bibinfo {author} {\bibfnamefont {E.}~\bibnamefont
  {Abbas}} \emph {et~al.} (\bibinfo {collaboration} {ALICE}),\ }\href {\doibase
  10.1140/epjc/s10052-013-2496-5} {\bibfield  {journal} {\bibinfo  {journal}
  {Eur. Phys. J. C}\ }\textbf {\bibinfo {volume} {73}},\ \bibinfo {pages}
  {2496} (\bibinfo {year} {2013})},\ \Eprint {http://arxiv.org/abs/1305.1562}
  {arXiv:1305.1562 [nucl-ex]} \BibitemShut {NoStop}%
\bibitem [{\citenamefont {Chernodub}(2016)}]{Chernodub:2015gxa}%
  \BibitemOpen
  \bibfield  {author} {\bibinfo {author} {\bibfnamefont {M.~N.}\ \bibnamefont
  {Chernodub}},\ }\href {\doibase 10.1007/JHEP01(2016)100} {\bibfield
  {journal} {\bibinfo  {journal} {JHEP}\ }\textbf {\bibinfo {volume} {01}},\
  \bibinfo {pages} {100} (\bibinfo {year} {2016})},\ \Eprint
  {http://arxiv.org/abs/1509.01245} {arXiv:1509.01245 [hep-th]} \BibitemShut
  {NoStop}%
\bibitem [{\citenamefont {Frenklakh}(2016)}]{Frenklakh:2016izv}%
  \BibitemOpen
  \bibfield  {author} {\bibinfo {author} {\bibfnamefont {D.}~\bibnamefont
  {Frenklakh}},\ }\href {\doibase 10.1103/PhysRevD.94.116010} {\bibfield
  {journal} {\bibinfo  {journal} {Phys. Rev. D}\ }\textbf {\bibinfo {volume}
  {94}},\ \bibinfo {pages} {116010} (\bibinfo {year} {2016})},\ \Eprint
  {http://arxiv.org/abs/1603.08971} {arXiv:1603.08971 [hep-th]} \BibitemShut
  {NoStop}%
\bibitem [{\citenamefont {Frenklakh}\ and\ \citenamefont
  {Gorsky}(2017)}]{Frenklakh:2017grl}%
  \BibitemOpen
  \bibfield  {author} {\bibinfo {author} {\bibfnamefont {D.}~\bibnamefont
  {Frenklakh}}\ and\ \bibinfo {author} {\bibfnamefont {A.}~\bibnamefont
  {Gorsky}},\ }\href {\doibase 10.1103/PhysRevD.96.034003} {\bibfield
  {journal} {\bibinfo  {journal} {Phys. Rev. D}\ }\textbf {\bibinfo {volume}
  {96}},\ \bibinfo {pages} {034003} (\bibinfo {year} {2017})},\ \Eprint
  {http://arxiv.org/abs/1703.02516} {arXiv:1703.02516 [hep-th]} \BibitemShut
  {NoStop}%
\bibitem [{\citenamefont {Ding}\ \emph {et~al.}(2023)\citenamefont {Ding},
  \citenamefont {Gu}, \citenamefont {Kumar}, \citenamefont {Li},\ and\
  \citenamefont {Liu}}]{Ding:2023bft}%
  \BibitemOpen
  \bibfield  {author} {\bibinfo {author} {\bibfnamefont {H.-T.}\ \bibnamefont
  {Ding}}, \bibinfo {author} {\bibfnamefont {J.-B.}\ \bibnamefont {Gu}},
  \bibinfo {author} {\bibfnamefont {A.}~\bibnamefont {Kumar}}, \bibinfo
  {author} {\bibfnamefont {S.-T.}\ \bibnamefont {Li}}, \ and\ \bibinfo {author}
  {\bibfnamefont {J.-H.}\ \bibnamefont {Liu}},\ }\href@noop {} {\  (\bibinfo
  {year} {2023})},\ \Eprint {http://arxiv.org/abs/2312.08860} {arXiv:2312.08860
  [hep-lat]} \BibitemShut {NoStop}%
\end{thebibliography}%

\end{document}